\documentclass[11pt, a4paper]{article}
\usepackage[labelfont=bf,textfont=normal]{caption}
 \usepackage[font=small, format=hang, margin={1cm, 1cm}]{caption}
\usepackage[a4paper, left=3cm,right=3cm]{geometry}
\usepackage{amsfonts} 
\usepackage{amsmath} 
\usepackage{mathrsfs}  %
\usepackage{amssymb}
\usepackage{mathtools}
\usepackage{relsize}
\usepackage{setspace}   
\usepackage{color}  
\usepackage{wasysym}
\usepackage{booktabs}
\usepackage[utf8,applemac]{inputenc} 
\usepackage{tensor}
\usepackage{cite}
\usepackage{tikz}
\usetikzlibrary{calc, arrows.meta}
\usepackage{graphicx}
\usepackage{subfig}
\usepackage{caption}
\usepackage[normalem]{ulem}
\usepackage{hyperref}
\usepackage{epstopdf}
\usepackage{comment}

\numberwithin{equation}{section}
\usepackage{dcolumn}
\usepackage{bm}
\usepackage[toc,page]{appendix}

\newcommand{\beq}{\begin{equation}}
\newcommand{\eeq}{\end{equation}}
\newcommand{\beqn}{\begin{eqnarray}}
\newcommand{\eeqn}{\end{eqnarray}}
\newcommand{\pa}{\partial}

\usepackage{cancel}

\newcommand{\jf}{\varphi}
\newcommand{\pu}{\partial_u}
\newcommand{\mM}{\mathcal {M}}
\newcommand{\mJ}{\mathcal {J}}

\def\ndelta{\delta\hspace{-0.50em}\slash\hspace{-0.05em} }

\newcommand\cnote[1]{\textcolor{red}{\bf [C:\,#1]}}
\definecolor{bluegreen}{RGB}{0,102,102}

\usepackage{pict2e}
\makeatletter
\newcommand{\loplus}{\mathbin{\mathpalette\dog@lsemi{+}}}
\newcommand{\dog@lsemi}[2]{\dog@semi{#1}{#2}{270,90}}
\newcommand{\dog@semi}[3]{%
  \begingroup
  \sbox\z@{$\m@th#1#2$}%
  \setlength{\unitlength}{\dimexpr\ht\z@+\dp\z@\relax}%
  \makebox[\wd\z@]{\raisebox{-\dp\z@}{%
    \begin{picture}(1,1)
    \linethickness{\variable@rule{#1}}
    \roundcap
    \put(0.5,0.5){\makebox(0,0){\raisebox{\dp\z@}{$\m@th#1#2$}}}
    \put(0.5,0.5){\arc[#3]{0.5}}
    \end{picture}%
  }}%
  \endgroup
}
\newcommand{\variable@rule}[1]{%
  \fontdimen8  
  \ifx#1\displaystyle\textfont3\else
    \ifx#1\textstyle\textfont3\else
      \ifx#1\scriptstyle\scriptfont3\else
        \scriptscriptfont3\relax
  \fi\fi\fi
}
\makeatother

\hypersetup{
    colorlinks,%
    citecolor=blue,%
    filecolor=blue,%
    linkcolor=blue,%
   urlcolor=blue,
   linktoc=page
}

\begin{document}


\setcounter{tocdepth}{2}

\begin{titlepage}

$ $
\vspace{20pt}

\begin{center}
{ \LARGE{\bf{Conservation and Integrability in Lower-Dimensional Gravity}}} 

 \vspace{5mm}

\vspace{2mm} 
\centerline{\large{\bf{Romain Ruzziconi\footnote{e-mail: romain.ruzziconi@tuwien.ac.at},  C\'eline Zwikel\footnote{e-mail: celine.zwikel@tuwien.ac.at}}}}

\vspace{2mm}
\normalsize
\bigskip\medskip
\textit{Institute for Theoretical Physics, TU Wien,\\
Wiedner Hauptstrasse 8, A-1040 Vienna, Austria\\
\vspace{2mm}
}
\vspace{25mm}

\begin{abstract}
    
    We address the questions of conservation and integrability of the charges in two and three-dimensional gravity theories at infinity. The analysis is performed in a framework that allows us to treat simultaneously asymptotically locally AdS and asymptotically locally flat spacetimes. In two dimensions, we start from a general class of models that includes JT and CGHS dilaton gravity theories, while in three dimensions, we work in Einstein gravity. In both cases, we construct the phase space and renormalize the divergences arising in the symplectic structure through a holographic renormalization procedure. We show that the charge expressions are generically finite, not conserved but can be made integrable by a field-dependent redefinition of the asymptotic symmetry parameters. 
    


\end{abstract}


\end{center}

\end{titlepage}

\newpage
\tableofcontents

\newpage
\section*{Introduction}
\addcontentsline{toc}{section}{Introduction}

In four-dimensional gravity, the analysis of asymptotically flat spacetimes at null infinity \cite{Bondi:1962px , Sachs:1962wk} has led to the Bondi mass loss formula, which states that the mass of the system decreases in time due to the emission of gravitational waves. This theoretical result was one of the striking arguments to prove the existence of the gravitational waves at a non-linear level of the theory. The non-conservation of charges is therefore an important ingredient to describe the dynamics of the system \cite{1977asst.conf....1G , Ashtekar:2014zsa, Held:1980gc}. When applying covariant phase space methods \cite{Crnkovic:1986ex , Lee:1990nz , Iyer:1994ys , Wald:1993nt , Barnich:2001jy , Barnich:2003xg , Barnich:2007bf} to derive the gravitational charges from first principles, non-conservation has been observed to be related to non-integrability of the charges \cite{Wald:1999wa , Barnich:2011mi,  Barnich:2013axa}.

This non-integrability is often considered as an unpleasant property since it implies that the finite charge expressions depend on the particular path that one chooses to integrate on the solution space, which is a typical feature of a dissipative system. Several prescriptions have been proposed to isolate meaningful integrable parts in the charge expressions \cite{Barnich:2011ty , Wald:1999wa , Flanagan:2015pxa , Chandrasekaran:2018aop , Compere:2018ylh , Haco:2018ske ,   Chandrasekaran:2020wwn, Chen:2020bhg}. These procedures require additional inputs in the theory, which rely on the context and the specific motivations. An alternative approach is to keep the full non-integrable expressions and try to make sense of them. An important technical result going into this direction is the Barnich-Troessaert bracket \cite{Barnich:2011mi} that allows one to derive mathematically consistent charge algebras for non-integrable charges. This bracket has then been used in many different contexts \cite{Barnich:2013axa , Donnay:2016ejv , Compere:2018ylh , Barnich:2019vzx , Adami:2020amw , Compere:2020lrt , Chandrasekaran:2020wwn , Fiorucci:2020xto} and the associated charge algebras have been shown to be physically extremely relevant since they contain all the information about the flux-balance laws of the theory \cite{Wald:1999wa  , Compere:2019gft ,Godazgar:2018vmm }. 

Recently, progress has been made in understanding the relation between non-conservation and non-integrability of the charges. In \cite{Adami:2020ugu}, it was proposed that no ``genuine'' flux passing through the boundary is equivalent to the existence of a particular slicing of the phase space for which the charges are integrable. This conjecture has been shown to hold for generic hypersurfaces in the bulk for topological theories \cite{Adami:2020ugu} and some preliminary results suggest that it is also true in four-dimensional gravity  \cite{progress-1}. A natural but non-trivial question that we explore in this work is whether this conjecture is also applicable for asymptotic boundaries.

A widely used gauge to study asymptotic boundaries is the Bondi gauge \cite{Bondi:1962px , Sachs:1962wk , Tamburino:1966zz , Barnich:2010eb}. Indeed, it is particularly well-adapted to investigate the interplay between radiation and symmetries \cite{Strominger:2014pwa , Pasterski:2015tva , Flanagan:2015pxa , Campiglia:2014yka , Nichols:2018qac , Compere:2018ylh , Tahura:2020vsa , Compere:2019gft , Blanchet:2020ngx}. Furthermore, it allows us to consider simultaneously asymptotically locally flat spacetimes exhibiting null boundaries and asymptotically locally AdS spacetimes with timelike boundaries \cite{Barnich:2012aw , Poole:2018koa , Compere:2020lrt , Ciambelli:2020eba , Ciambelli:2020ftk , Godet:2020xpk , Ruzziconi:2020cjt}. The analyses of these two types of asymptotics are related through a flat limit process \cite{Barnich:2012aw ,Ciambelli:2018wre , Compere:2019bua}.

To consider asymptotically locally AdS and flat spacetimes in Bondi gauge, one has to allow the boundary structure to fluctuate \cite{Campiglia:2014yka, Campiglia:2015yka ,
Poole:2018koa , Compere:2018ylh , Compere:2020lrt , Ciambelli:2020eba , Ciambelli:2020ftk}. These mild falloffs led to the proposal of an infinite-dimensional enhancement of the BMS group with smooth superrotations in the flat case \cite{Campiglia:2014yka, Campiglia:2015yka, Compere:2018ylh} and the discovery of BMS-like symmetries in presence of non-vanishing cosmological constant \cite{Compere:2019bua , Compere:2020lrt}. Fluctuations of the boundary structure are compatible with conformal compactification of the spacetime \cite{Penrose:1965am}, but involve some divergences at the level of the symplectic structure \cite{Compere:2018ylh , Flanagan:2019vbl , Compere:2020lrt}. While the holographic renormalization procedure is well understood in asymptotically locally AdS spacetimes written in Fefferman-Graham gauge \cite{deHaro:2000xn , Bianchi:2001kw,  Compere:2008us , Compere:2020lrt , Fiorucci:2020xto}, it is less clear how this works in Bondi gauge. One of the main objectives of this paper is to make one step further into that endeavour and investigate this question in lower-dimensional gravity. 

More precisely, in this work, we address the questions of conservation, integrability and renormalization of the charges in the framework of two-dimensional dilaton gravity and three-dimensional Einstein's gravity theories (see \textit{e.g.} \cite{brown1988lower, Grumiller:2002nm , Witten:1988hc}) in Bondi gauge. Indeed, these models share the common feature to have no local propagating degrees of freedom in the bulk of the spacetime. As such they provide a useful arena for investigating new concepts and ideas.

Analogously to the aforementioned higher-dimensional case, considering fluctuations of the boundary structure in lower-dimensional gravity theories unveils new asymptotic symmetries and the phase space analysis yields non-conserved and \textit{a priori} non-integrable charges at the asymptotic boundary. Since the theories are topological, the non-integrability and non-conservation are not due to the leak of gravitational waves through the spacetime boundary. Instead, these features can be seen as implied by the presence of external sources encoded in the fluctuations of the boundary structure \cite{Troessaert:2015nia , Fiorucci:2020xto  , Wieland:2020gno}. In this picture, the gravitational system is therefore seen as an open dissipative system. Freezing the fluctuations of the boundary structure, which amounts to turn off the sources, yields back a closed system. The precise nature of the external sources depends on the environment that one is considering. However, the analysis of the asymptotic structure does not require to provide a specific environment and one can directly work with the open system.

Analysing non-conservation, non-integrability and renormalization of the charges in two and three dimensions is definitely worthwhile since this sets the stage for similar investigations in higher dimensions. Beyond these technical considerations, the boundary conditions with fluctuating boundary structure that we consider may also have their own physical interest in lower-dimensional gravity theories. An example where they may be relevant appears in the recent analysis of the black hole information paradox to derive the Page curve from quantum gravity path integral arguments in two dimensions \cite{Almheiri:2019yqk , Almheiri:2019qdq} (see also \cite{Almheiri:2020cfm , Raju:2020smc}). In this context, it has been useful to couple the gravitational system with an environment so that the black hole can evaporate in AdS. Our considerations of fluctuating boundary conditions makes this construction explicit at the level of the asymptotic structure. Another context where the fluctuating boundary structure is relevant appears when considering brane worlds interacting with ambient higher-dimensional spacetimes \cite{Randall:1999ee , Randall:1999vf , Geng:2020fxl}. This picture naturally yields fluctuations of the boundary metric and induced quantum gravity on the boundary \cite{Compere:2008us}. Finally, let us mention that a specific example which reinforces the physical relevance of this type of relaxed boundary conditions has been investigated recently in asymptotically locally AdS$_3$ spacetimes, where the non-conservation of the charges has been interpreted as an anomalous Ward-Takahashi identity in the dual theory \cite{Alessio:2020ioh}. 

At a technical level, our motivation to work in the second order metric formalism is twofold: it has the great advantage to be very intuitive when imposing boundary conditions on the metric components and the results can be extended to the treatment of higher-dimensional cases without conceptual obstructions. For related works in other formalisms see \textit{e.g.} \cite{Grumiller:2016pqb,Grumiller:2017sjh,Grumiller:2017qao, Barnich:2019vzx,Barnich:2016rwk, Barnich:2020ciy, Godazgar_2019,Godazgar:2020gqd,Godazgar:2020kqd, Oliveri:2019gvm,Oliveri:2020xls,Freidel:2020xyx,Freidel:2020svx,Freidel:2020ayo ,Geiller:2020edh,Geiller:2020okp}. Note that it is expected that different formalisms lead to different symmetry algebras \cite{Freidel:2020xyx}.

\subsubsection*{Summary of the results}
Our analysis provides some new results concerning the quest of the most general boundary conditions in two dimensions for a generic class of models that contains JT \cite{Jackiw:1984je ,Teitelboim:1983ux} and CGHS \cite{Callan:1992rs} gravity theories. Indeed, as suggested in \cite{Grumiller:2017sjh , Grumiller:2016pqb} (see also \cite{Ciambelli:2020eba , Ciambelli:2020ftk , Adami:2020amw}), fixing the gauge completely in the analysis of asymptotics seems to eliminate some potentially interesting asymptotic symmetries and associated towers of charges. In this work, we construct the maximal asymptotic symmetry algebra that one can obtain in two-dimensional gravity theories by imposing only partial gauge fixing on the components of the metric and very mild falloffs. After a renormalization procedure of the on-shell action and the symplectic structure involving covariant counter-terms \cite{deHaro:2000vlm , Compere:2008us , Bergamin:2007sm , Grumiller:2007ju , Cvetic:2016eiv}, we obtain finite charge expressions that we render integrable through a field-dependent redefinition of the symmetry parameters \cite{Barnich:2007bf , Compere:2017knf , Grumiller:2019fmp , Adami:2020ugu , Alessio:2020ioh, Ciambelli:2020shy}. The asymptotic symmetry algebra is given by the direct sum of three abelian Lie algebras of smooth functions $C_\infty(\mathbb{R})\oplus C_\infty(\mathbb{R}) \oplus C_\infty(\mathbb{R})$. The associated charge algebra is shown to be centrally extended, exhibiting a Heisenberg subalgebra. Finally, a notion of flat limit is discussed between JT and CGHS gravity theories to relate the analysis performed in the different types of asymptotics \cite{Cangemi:1992bj , Verlinde:1991rf} (see also \cite{Afshar:2019axx , Grumiller:2020elf ,Gomis:2020wxp} and references therein).

In three dimensions, the rigidity of the Bondi gauge fixing does not allow us to pretend for the most general boundary conditions \cite{Ciambelli:2020eba , Ciambelli:2020ftk}. However, our analysis investigates new boundary conditions that encompass those considered previously in the literature (see \textit{e.g.} \cite{Brown:1986nw , Barnich:2010eb , Barnich:2012aw}). Furthermore, this set-up is sufficient to illustrate our techniques on renormalization and integrability of non-conserved charges.  After performing the holographic renormalization of the action and the symplectic structure in Bondi gauge using covariant counter-terms \cite{deHaro:2000vlm , Compere:2008us , Bergamin:2007sm , Detournay:2014fva , Fiorucci:2020xto} and corner terms \cite{Compere:2020lrt,Detournay:2014fva}, we obtain the gravitational charges associated with the solution space derived in \cite{Ciambelli:2020eba , Ciambelli:2020ftk} that includes fluctuating boundary structure. Again, we find the appropriate field-dependent redefinition of the parameters to render the charges integrable  \cite{Barnich:2007bf , Compere:2017knf , Grumiller:2019fmp , Adami:2020ugu , Alessio:2020ioh , Ciambelli:2020shy}. In asymptotically locally AdS$_3$ spacetimes (respectively asymptotically locally flat spacetimes), the asymptotic symmetry algebra is given by a Lie algebroid \cite{Crainic} with a one-dimensional base space parametrized by the time $u$ on the boundary and with a Diff$(S^1)\oplus$Diff$(S^1)$ (respectively BMS$_3$) algebra at each value of $u$. The charge algebra involves a central extension that reduces to the standard expressions when fixing the boundary structure: in asymptotically AdS$_3$ spacetimes, it reduces to the Brown-Henneaux central extension \cite{Brown:1986nw}, while in asymptotically flat spacetimes, it reproduces the BMS$_3$ central extension \cite{Barnich:2006av}.

\subsubsection*{Organization of the paper}
The paper is organized as follows. In section \ref{Integrability and conservation in the covariant phase space formalism}, we review the notions of non-conservation and non-integrability in the covariant phase space formalism, and mention the implication of the non-conservation at the level of the variational principle of the theory. In particular, we propose a refinement of the variational principle to accommodate open systems. In section \ref{sec:Gravity in two dimensions}, we present our results in two dimensions. More specifically, we apply the covariant phase space methods on a very general class of two-dimensional dilaton gravity models that includes JT and CGHS gravity theories. Then, we specify our analysis to the Bondi gauge with linear dilaton, keeping an arbitrary potential. We discuss the solution space (see also appendix \ref{app2d}), renormalize the action principle and the symplectic structure, derive the corresponding finite surface charges associated with the asymptotic symmetries, compute the charge algebra and discuss the flat limit. In section \ref{sec:Gravity in three dimensions}, we repeat this analysis for three-dimensional gravity theory. We conclude the discussion in the last section by providing further comments on the results.

\section{Conservation and integrability in the covariant phase space formalism}
\label{Integrability and conservation in the covariant phase space formalism}

In this section, we review the notions of non-conservation and non-integrability of charges. In particular, we explain the implication of non-conservation on the variational principle. We also review how the notion of integrability can be formulated as a Pfaff problem, which makes the field-dependent redefinitions of the symmetry parameters very natural.

\subsection{Conservation and variational principle}
\label{sec:Conservation and variational principle}

In the covariant phase space formalism \cite{Crnkovic:1986ex , Lee:1990nz , Iyer:1994ys , Wald:1993nt , Barnich:2001jy , Barnich:2003xg , Barnich:2007bf, Margalef-Bentabol:2020teu}, the infinitesimal charges associated with the asymptotic symmetry parameters $\xi$ are computed by integrating some co-dimension $2$ forms\footnote{Our convention for the components of differential forms is the following: a co-dimension $p$ form (or a $(n-p)$-form, where $n$ is the spacetime dimension) $\alpha$ is written as $
\alpha =  \alpha^{\mu_1 \ldots \mu_p} (d^{n-p} x)_{\mu_1 \ldots \mu_p}$ with $(d^{n-p} x)_{\mu_1 \ldots \mu_p} = \frac{1}{p!(n-p)!} \epsilon_{\mu_1 \ldots \mu_p \nu_1 \ldots \nu_{n-p}} dx^{\nu_1} \wedge \ldots \wedge dx^{\nu_{n-p}}.
$} $k_\xi [\phi; \delta \phi]$ on a co-dimension $2$ surface at infinity\footnote{In two dimensions, $S_\infty$ is a point on a boundary. In three dimensions, it corresponds to a circle $S^1$ on the boundary.} $S_\infty$ as
\begin{equation}
   \ndelta Q_\xi[\phi] = \int_{S_\infty} k_\xi [\phi; \delta \phi]
   \label{infinitesima charge definition}
\end{equation} where $\phi$ denotes the dynamical fields of the theory. As we will see later in the text, one can either use the Barnich-Brandt \cite{Barnich:2001jy , Barnich:2003xg , Barnich:2007bf} or the Iyer-Wald \cite{Lee:1990nz , Iyer:1994ys , Wald:1993nt ,Wald:1999wa} methods to construct this co-dimension $2$ form. For the purpose of this discussion, we take the later point of view that we briefly review now. 

Starting from a Lagrangian field theory $L=L[\phi]$, the presymplectic potential $\Theta[\phi; \delta \phi]$ is obtained by taking a variation on the field space
\begin{equation}
    \delta L[\phi] = \frac{\delta L[\phi]}{\delta \phi} \delta \phi + d \Theta [\phi; \delta \phi]\,.
    \label{variation lagrange theta}
\end{equation} This presymplectic potential is defined up to a $\delta$-exact co-dimension $1$ form $\delta A[\phi]$ and up to a $d$-exact co-dimension $1$ form $d Y[\phi ; \delta \phi]$, namely
\begin{equation}
    \Theta [\phi; \delta \phi] \to \Theta [\phi; \delta \phi] + \delta A[\phi] +  d Y[\phi ; \delta \phi]\,.
    \label{ambiguity in theta}
\end{equation} The $\delta$-exact ambiguity is coming from the freedom to add boundary terms to the bulk Lagrangian $L[\phi] \to L[\phi] + d A[\phi]$. It can be fixed by prescribing the action principle $S$. The $d$-exact term in \eqref{ambiguity in theta} is due to the fact that $d^2 = 0$ so that \eqref{variation lagrange theta} remains unaffected by adding this term\footnote{See for \textit{e.g.}\cite{Freidel:2020xyx} for a prescription to fix the $d$-exact ambiguity. }. The presymplectic current is then defined through
\begin{equation}
    \omega [\phi ; \delta_1 \phi, \delta_2 \phi] = \delta_2 \Theta[\phi; \delta_1 \phi] - \delta_1 \Theta[\phi ;\delta_2 \phi]\,.
    \label{IW presumplecticc}
\end{equation} 
Notice that from \eqref{ambiguity in theta}, it is defined up to the ambiguity
\begin{equation}
    \omega [\phi ; \delta_1 \phi, \delta_2 \phi] \to \omega [\phi ; \delta_1 \phi, \delta_2 \phi] + \delta_2 d Y[\phi; \delta_1 \phi] - \delta_1 d Y[\phi; \delta_2 \phi]\,.
     \label{ambiguity current}
\end{equation} In particular, the $\delta$-exact ambiguity in \eqref{ambiguity in theta} does not influence the presymplic current, neither the charges. The presymplectic potential is related to the on-shell variation of the action through
\begin{equation}
    \delta S = \int_{\mathscr{I}} \Theta [\phi; \delta \phi]
    \label{delta S}
\end{equation} where we consider only timelike or null spacetime boundaries $\mathscr{I}$. 

The Iyer-Wald co-dimension $2$-form $k_\xi[\phi; \delta \phi]$ appearing in \eqref{infinitesima charge definition} can be defined on-shell through 
\begin{equation}
    d k_\xi [\phi; \delta \phi] = \omega [\phi; \delta_\xi \phi, \delta \phi]
    \label{breaking nin conservation}
\end{equation} (see sections \ref{sec:Phase space of dilaton gravity models} and \ref{Phase space of three-dimensional gravity} for explicit constructions). It is determined up to a $d$-exact co-dimension $2$ form $k_\xi [\phi; \delta \phi] \to k_\xi [\phi; \delta \phi] + d M_\xi [\phi; \delta \phi]$. This ambiguity does not play any role when integrating on a compact co-dimension $2$ surface as in \eqref{infinitesima charge definition}. Of course, the ambiguity in the presymplectic current \eqref{ambiguity current} brings some non-trivial modifications at the level of the co-dimension $2$ form $k_\xi [\phi; \delta \phi]$. In particular, this ambiguity will be used in the following to renormalize the divergences arising in the symplectic structure.

The non-conservation of the infinitesimal charge \eqref{infinitesima charge definition} is completely controlled by the breaking in the closure of the co-dimension $2$ form $k_\xi[\phi; \delta \phi]$ in \eqref{breaking nin conservation}. In particular, if the pull-back of the on-shell presymplectic current vanishes on the spacetime boundary $\mathscr{I}$, \textit{i.e.} 
\begin{equation}
    \omega [\phi; \delta_1 \phi, \delta_2 \phi]|_{\mathscr{I}} = 0\,,
    \label{conservative boundary conditions}
\end{equation}  the infinitesimal charge will be  conserved in time. Notice that in this case, the pull-back of the presymplectlic potential is necessarily a $\delta$-exact term, \textit{i.e.} $\Theta[\phi ; \delta \phi] |_\mathscr{I} = \delta B[\phi]$. This implies that one can add a boundary term to the action, 
\begin{equation}
    S \to S - \int_\mathscr{I} B [\phi]
\end{equation} so that the variational principle is stationary on-shell, \textit{i.e.} $\delta S =0$ (see \eqref{delta S}). In asymptotically AdS spacetimes, following the terminology of \cite{Fiorucci:2020xto}, boundary conditions that fulfil the requirement \eqref{conservative boundary conditions} are called conservative boundary conditions. The archetype of those is given by the Dirichlet boundary conditions which freeze completely the boundary metric.

As argued in \cite{Fiorucci:2020xto}, considering leaky boundary conditions, namely boundary conditions for which 
\begin{equation}
    \omega [\phi; \delta_1 \phi, \delta_2 \phi]|_{\mathscr{I}} \neq 0
    \label{leaky bc}
\end{equation} is appealing. While this type of boundary conditions is mandatory when considering radiative asymptotically flat spacetimes at null infinity in four dimensions, their analysis in asymptotically AdS spacetimes is very recent and their implication on the AdS/CFT is still to be uncovered. Importantly, for leaky boundary conditions, the presymplectic potential is not $\delta$-exact and one cannot obtain a well-defined variational principle satisfying $\delta S = 0$ on-shell. We therefore impose refined criteria to (partially) fix the boundary terms of the action:
\begin{itemize}
     \item The action principle is well-defined when restricting to Dirichlet boundary conditions, namely we recover the standard result $\delta S =0$ on-shell\footnote{Notice that the first requirement echoes the Dirichlet flux condition imposed in \cite{Chandrasekaran:2020wwn} when considering non-vanishing flux in the phase space analysis.}. This allows us to interpret the fluctuations of the boundary structure as external sources responsible for the non-conservation of the infinitesimal charges.
     \item The action is finite on-shell. 
     Indeed, the Euclidean version of the on-shell action is associated to the free energy of the system in the canonical ensemble\footnote{However, see \textit{e.g.} \cite{Grumiller:2007ju , Berkovits:2001tg, Gukov:2003yp , Davis:2004xb} for examples where the requirement of finite free energy has to be refined.}. This finiteness requirement will allow us to prescribe covariant counter-terms to renormalize the symplectic structure.
\end{itemize}
Moreover, in this work, we require that the action admits a well-defined flat limit. This requirement only makes sense when one is considering a solution space that admits a well-defined flat limit, which is the case in Bondi gauge. 

We believe that these three requirements are the natural criteria to impose in presence of leaky boundary conditions in order to prescribe meaningful boundary terms for the action. We will illustrate them in the explicit examples considered below and perform the appropriate holographic renormalization (see sections \ref{sec:Holographic renormalization 2d} and \ref{Holographic renormalization in Bondi gauge}). 

Notice that leaky boundary conditions imply that we are dealing with an open gravitational system where the external sources are encoded in the variations of the boundary structure. In other words, the fluctuations of the boundary structure can be interpreted as boundary degrees of freedom that couple with the gravitational system. The fact that the action is not stationary on solutions arises because we do not include the environment into the analysis. Indeed, the investigation of asymptotics does not require to know the precise nature of this environment. Including the latter into the analysis would impose some specific dynamics for the sources and restore the stationarity of the action on the solutions\footnote{For instance, in the context of electromagnetism at spatial infinity, boundary degrees of freedom with constrained dynamics are introduced and lead to a well-defined action principle \cite{Henneaux:2018gfi , Henneaux:2018hdj}.}.  A typical example would be to take the boundary at finite distance and interpret it as a brane world in AdS with induced gravity \cite{Randall:1999ee , Randall:1999vf}.

\subsection{Integrability}
\label{sec:Integrability}

Let us now discuss the notion of integrability of the charges. The infinitesimal charge \eqref{infinitesima charge definition} is a $1$-form on the solution space. It can be integrated on a path $\gamma$ in the solution space to obtain the surface charge expression
\begin{equation}
    Q_\xi [\phi] = \int_\gamma \ndelta Q_\xi[\phi] + N_\xi
    \label{integration on solution space}
\end{equation} where $N_\xi$ is the value of the charge at the reference solution. This integration on the solution space is path-independent if and only if the infinitesimal charge is $\delta$-exact, \textit{i.e.} $\ndelta Q_\xi[\phi] \equiv \delta Q_\xi[\phi]$. If the infinitesimal charge is $\delta$-exact, we say that it is integrable. If this is not the case, we say that it is non-integrable.

As explained in \cite{Barnich:2007bf}, given an infinitesimal charge expression, the question whether or not it is possible to find an integrable charge by performing a field-dependent redefinition of the parameters can be addressed more precisely as a Pfaff system on the solution space governed by the Frobenius theorem\footnote{See \textit{e.g.} \cite{delphenich2017role} for a relevant discussion on Pfaff systems in a physical context.}. Let $x$ be the spacetime coordinates and $\phi = \phi(x;a)$ the solutions of the equations of motion parametrized by $a = (a^A)$, $A=1, \ldots , p$. Let us also write $\xi = \xi (x; a, b)$ the asymptotic symmetry generators involving the solution space parameters $a$ and depending linearly on symmetry parameters $b = (b^i)$, $i= 1 \ldots q$. The generators $e_i(x,a) = \frac{\partial}{\partial b^i} \xi (x; a, b)$, $i=1, \ldots, q$, form a basis of the Lie algebra at a given point in the solution space\footnote{A structure of Lie algebroid emerges naturally in the context of gauge symmetries and asymptotic symmetries \cite{Crainic ,Barnich:2010xq , Barnich:2017ubf}.}. Now, we consider the set of $1$-forms on the solution space $\theta_i [a, \delta a] = \ndelta Q_{e_i}[\phi (x;a); \delta \phi (x; a)]$. The question of integrability can be formulated as follows: is it possible to find a field-dependent invertible $q \times q$ matrix $S^i_j (a)$ such that
\begin{equation}
    \delta Q_{f_j} [\phi (x;a); \delta \phi (x; a)] = S^i_j (a) \theta_i [a, \delta a] = \ndelta Q_{ S^i_j (a) e_i}[\phi (x;a); \delta \phi (x; a)]
    \label{field depdent redef abstract}
\end{equation} where $f_j = S^i_j (a) e_i$? If the answer is yes, then the charges are integrable and the integration of the expressions $\delta Q_{f_j}[\phi; \delta \phi]$ on the solution as in \eqref{integration on solution space} is path independent. 

While the problem is well posed, it is not clear to know \textit{a priori} whether a generic system is integrable or not. Nevertheless, in the absence of propagating degrees of freedom passing through the boundary, it is physically expected that a slicing of the phase space for which the charges are integrable can be found \cite{Adami:2020ugu}.
In equations \eqref{new parameters} and \eqref{redefinition para 3d} below, we will see two explicit examples of field-dependent redefinition as in \eqref{field depdent redef abstract} (see also \cite{ Compere:2017knf , Grumiller:2019fmp ,  Alessio:2020ioh, Ciambelli:2020shy} for other examples).

Another interesting aspect is that if there exists an integrable slicing of the phase space, then it is not unique. In general there is an infinite number of slicings preserving the integrability of the charges. See \cite{Adami:2020ugu} for an explicit discussion. 

\section{Gravity in two dimensions}
\label{sec:Gravity in two dimensions}

We now apply the general framework presented in section \ref{Integrability and conservation in the covariant phase space formalism} to the case of two-dimensional dilaton gravity.

\subsection{Phase space of dilaton gravity models}
\label{sec:Phase space of dilaton gravity models}

In two-dimensional dilaton gravity, the dynamical fields of the theory are the metric $g_{\mu\nu}$ and the dilaton scalar field $X$. Writing $\phi= (g_{\mu\nu} , X)$ with $(x^\mu) =(x^0, x^1)$, we consider the general class of dilaton gravity models
\begin{equation} 
L_{DGT}[\phi] = \frac{\sqrt{-g}}{16 \pi G} [RX - U(X) (\nabla X)^2 - 2 V(X) ]
\label{DGT}
\end{equation} where $U(X)$ and $V(X)$ are potentials that are functions of the dilaton field $X$ (see \textit{e.g.} \cite{Grumiller:2002nm} for a review). This contains an important class of two-dimensional dilaton gravity theories, including JT gravity \cite{Jackiw:1984je ,Teitelboim:1983ux}
\begin{equation}
L_{JT}[\phi] =  \frac{\sqrt{-g}}{16\pi G} X[R - 2 \Lambda ], \qquad U(X) = 0, \, V(X) = \Lambda X
\label{JT model}
\end{equation} and CGHS model \cite{Callan:1992rs}
\begin{equation}
L_{CGHS}[\phi] =  \frac{\sqrt{-g}}{16\pi G} [X R - 2 \lambda], \qquad U(X) = 0, \, V(X) = \lambda 
\label{CGHS model}
\end{equation} where $\Lambda$ and $\lambda$ are constants whose relation in the flat limit will be discussed in section \ref{sec: Flat limit 2d}.

Taking an infinitesimal variation of the Lagrangian density \eqref{DGT}, and integrating by parts to keep track of the boundary terms, we have
\begin{equation}
\delta L_{DGT}[\phi] = \frac{\delta L_{DGT}}{\delta g_{\mu \nu}} \delta g_{\mu\nu} + \frac{\delta L_{DGT}}{\delta X} \delta X + \partial_\mu \Theta^\mu_{DGT} [\phi;\delta \phi] 
\end{equation} where the Euler-Lagrange derivatives are explicitly given by
\begin{equation}
    \begin{split}
\label{GMeomG}
\frac{\delta L_{DGT}}{\delta g_{\mu \nu}} &= \frac{\sqrt{-g}}{16 \pi G} \left[\nabla^\mu \nabla^\nu X - g^{\mu\nu} \nabla^2 X + (\nabla^\mu X) (\nabla^\nu X) U - \frac{1}{2} g^{\mu\nu} (\nabla X)^2 U - g^{\mu\nu} V\right] , \\
\frac{\delta L_{DGT}}{\delta X} &= \frac{\sqrt{-g}}{16 \pi G} [R + U' (\nabla X)^2 + 2 U \nabla^2 X - 2 V']
\end{split}
\end{equation} and the canonical presymplecic potential\footnote{We use the terminology ``canonical presymplectic potential" to refer to the presymplectic potential obtained directly by integration by parts from the Lagrangian, without adding any ambiguity appearing in \eqref{ambiguity in theta}.} reads as 
\begin{align}
\Theta_{DGT}^\mu [ \phi; \delta \phi ] &=   X {{\Theta}}_{EH}^\mu [g; \delta g] + \frac{\sqrt{-g}}{16\pi G} [- (\delta g)^{\mu \nu} \nabla_\nu X + (\delta g)^\nu_\nu \nabla^\mu X     - 2 \delta X (\nabla^\mu X) U ] \label{presymplectic}
\end{align}  where ${\Theta}_{EH}^\mu [g; \delta g]= \frac{\sqrt{-g}}{16\pi G} [ \nabla_\nu (\delta g)^{\mu\nu} - \nabla^\mu (\delta g)^\nu_\nu]$. In these expressions, the variation of the metric is defined with lower indices and $(\delta g)^{\mu\nu} = g^{\mu \alpha} g^{\nu \beta} \delta g_{\alpha \beta}$, $(\delta g)^\nu_\nu = g^{\alpha\nu} \delta g_{\alpha \nu}$. 

The theory \eqref{DGT} is invariant under diffeomorphisms that act on the dynamical fields as
\begin{equation}
    \delta_\xi g_{\mu\nu} = 2\nabla_{(\mu} \xi_{\nu)} , \qquad \delta_\xi X = \xi^\mu \nabla_\mu X 
\end{equation} where $\xi$ is the generator of infinitesimal diffeomorphisms.

Let us now derive the co-dimension $2$ forms (which are actually $0$-forms in two dimensions) associated with \eqref{DGT} that contain the information about the charges of the theory. We first apply the Barnich-Brandt procedure \cite{Barnich:2001jy , Barnich:2003xg , Barnich:2007bf} and then relate the results to the Iyer-Wald construction \cite{Lee:1990nz , Iyer:1994ys , Wald:1993nt ,Wald:1999wa}. The weakly-vanishing Noether current $S^\mu_\xi[\phi]$, is obtained through
\begin{equation}
\frac{\delta L_{DGT}}{\delta g_{\mu \nu}} \delta_\xi g_{\mu\nu} + \frac{\delta L_{DGT}}{\delta X} \delta_\xi X = \partial_\mu S^\mu_\xi[\phi],
\end{equation} where to obtain the right-hand side, we have integrated by parts in order to isolate the diffeomorphism parameters and used the Noether identities 
\begin{equation}
\frac{\delta L_{DGT}}{\delta g_{\mu \nu}} \partial_\rho g_{\mu \nu} + \frac{\delta L_{DGT}}{\delta X} \partial_\rho X - 2 \partial_\mu \left(  \frac{\delta L_{DGT}}{\delta g_{\mu \nu}} g_{\rho \nu} \right) = 0
\label{Noether identities}
\end{equation} which can be checked explicitly. The total derivative term gives the weakly-vanishing Noether current 
\begin{equation}
S^\mu_\xi [\phi] = 2\frac{\delta L_{DGT}}{\delta g_{\mu \nu}} g_{\rho \nu} \xi^\rho\,.
\label{wekly vanishing}
\end{equation} It has the property to vanish and to be conserved on-shell. Applying the homotopy operator on it yields the Barnich-Brandt co-dimension 2 form, which can be used to compute the charges associated with the diffeomorphism generator $\xi$, 
\begin{equation}
k^{\mu\nu}_{BB,\xi} [\phi; \delta \phi] = \frac{1}{2} \delta \phi^i \frac{\delta}{\delta \phi_\nu^i} S^\mu_\xi + \left( \frac{2}{3} \partial_\sigma\delta \phi^i-\frac13 \delta \phi^i\partial_\sigma\right) \frac{\delta}{\delta \phi^i_{\nu \sigma}} S^\mu_\xi - (\mu \leftrightarrow \nu) \label{homotopy operator}
\end{equation} where $(\phi^i) = (X, g_{\mu\nu})$. We have explicitly 
\begin{align} 
k_{BB,\xi}^{\mu\nu} [\phi; \delta \phi ]=& 
 \frac{\sqrt{-g}}{8\pi G}\left[ 2 U \xi^{[\nu} (\nabla^{\mu]}X) \delta X +2(\nabla^{[\mu}\delta X) \xi^{\nu]} - \delta X \nabla^{[\mu}\xi^{\nu]} 
 +\xi^{[\mu}(\delta g)^{\nu]}_\alpha \nabla^\alpha X \right] \,.
 \label{BB codimension 2}
\end{align} It is worthy to remark that the charge expression is independent of the potenital $V$ but depends on $U$. This co-dimension $2$ form is defined without any ambiguity since there is no possibility for total derivative terms in two dimensions. 

Now, the relation between Barnich-Brandt and Iyer-Wald procedures is controlled by the object
\begin{equation}
    E^{\mu\nu}[\phi; \delta_1 \phi, \delta_2 \phi] =  \frac{1}{16 \pi G} X (\delta_1 g)^\mu_\sigma  (\delta_2 g)^{\nu \sigma} - (1\leftrightarrow 2)
    \label{Link IW BB 2d}
\end{equation} so that
\begin{equation}
    k_{BB,\xi}^{\mu \nu}[\phi; \delta \phi] =  k_\xi^{\mu \nu}[\phi; \delta \phi] - E^{\mu\nu}[\phi; \delta_\xi \phi, \delta \phi]\,.
    \label{Link IW BB 2d v2}
\end{equation} We have the conservation law \eqref{breaking nin conservation},
\begin{equation}
    \partial_\nu k^{\mu\nu}_{\xi} [\phi; \delta \phi] = \omega^\mu_{DGT}[\phi;\delta_\xi \phi , \delta \phi]
    \label{conservation IW 2d}
\end{equation} where $\omega^\mu_{DGT}[g; \delta_1 g, \delta_2 g] = \delta_2 \Theta_{DGT}^\mu  [g; \delta_1 g] - \delta_1 \Theta_{DGT}^\mu[g; \delta_2 g] $ is the Iyer-Wald presymplectic current defined in \eqref{IW presumplecticc}. In the following, we will work in the Iyer-Wald approach that allows us to renormalize the symplectic structure using the ambiguities arising in \eqref{ambiguity in theta}.

\subsection{Linear dilaton Bondi gauge in two dimensions}
\label{Linear Dilaton Bondi gauge in two dimensions}

We now apply the formalism displayed in the previous section to study the asymptotic structure of the spacetime. For convenience, we set the kinetic potential $U(X)$ to zero in \eqref{DGT} but keep $V(X)$ arbitrary (in particular, this discussion includes JT \eqref{JT model} and CGHS \eqref{CGHS model} models). This choice eliminates the dependence in the potential in the expression of the charges \eqref{BB codimension 2} and therefore allows us to keep the discussion of the phase space very generic. We refer to appendix \ref{app2d} for a discussion of the solution space that includes an arbitrary $U(X)$. 

\subsubsection{Solution space}

Writing the coordinates as $(x^\mu) = (u,r)$, we start by imposing the following condition on the metric:
\begin{equation}
    g_{rr} = 0 \,.
    \label{gauge fixing}
\end{equation} We will refer to this condition as the Bondi (partial) gauge fixing on the metric, analogously to the terminology commonly used in higher dimensions \cite{Bondi:1962px , Sachs:1962wk , Barnich:2010eb}. Notice that we have only used one degree of freedom in the diffeomorphims among the two available in two dimensions. Therefore, this is only a partial gauge fixing (see \textit{e.g} \cite{Ruzziconi:2019pzd}). We write the metric as \begin{equation}
    ds^2 = 2 B(u,r) du^2 - 2 e^{A(u,r)} du dr\,.
    \label{Bondi metric 2d}
\end{equation} Furthermore, we will consider linear dilaton gravity solutions which are relevant for our analysis of the phase space since it will produce non-vanishing charges. Using the residual gauge diffeomorphisms preserving \eqref{gauge fixing}, we set
\begin{equation}
    X= e^{-Q_0(u)} r+ \jf_0 (u) \,.
    \label{Linear dilaton}
\end{equation}

Solving the equations of motion \eqref{GMeomG} in the linear dilaton Bondi gauge (conditions \eqref{Bondi metric 2d} and \eqref{Linear dilaton}), we obtain 
\begin{equation}
\begin{split}
    A &= A_0 (u) ,\\
    B &=e^{Q_0+A_0} B_0+e^{A_0} \partial_u Q_0 \, r +e^{2(A_0+Q_0)}\int^Xd Y V(Y)\,.
    \end{split}
\end{equation} 
The last equation of motion \eqref{app2devoleq} is conveniently written as
\begin{equation}
    \partial_u \mathcal{M} = 0\,,
    \label{2devoleq}
\end{equation} 
in terms of
\begin{equation}\label{2dCasimir}
    \mathcal{M}= e^{-(A_0+Q_0)} (B_0+ \partial_u \varphi_0)\,.
\end{equation}
It corresponds to a Casimir of the theory \cite{Klosch:1995fi , Grumiller:2002nm} and labels the different orbits for the action of the symmetry group on the solution space (see equation \eqref{VARIATIONS 2D} below). When considering the surface charges in section \ref{Surface charges in 2d}, $\mathcal{M}$ will be interpreted as the mass of the system.

In summary, the solution space is thus characterized by four functions of $u$: ($Q_0(u)$, $A_0(u)$, $\jf_0(u)$, $B_0(u)$) with the constraint \eqref{2devoleq} on the Casimir. 

\subsubsection{Residual symmetries}
The generators of the residual gauge diffeomorphisms preserving the linear dilaton Bondi gauge (equations \eqref{Bondi metric 2d} and \eqref{Linear dilaton}) are given by
\begin{equation}
    \xi= \epsilon (u)  \partial_u +( \chi (u) \, r+ \eta (u) )\partial_r
    \label{residual 2d}
\end{equation} where $\epsilon (u)$, $\chi (u)$ and $\eta (u)$ are arbitrary functions of $u$ that may be field-dependent. Using the modified Lie bracket\footnote{The modified Lie bracket is sometimes referred as the adjusted bracket (see \textit{e.g} \cite{Adami:2020amw , Adami:2020ugu}). The terminology should not be confused with the modified Barnich-Troessaert bracket at the level of the charge algebra \cite{Barnich:2011mi}.} \cite{Barnich:2010eb}
\begin{equation}
[\xi_1, \xi_2]_\star = [\xi_1 , \xi_2] - \delta_{\xi_1} \xi_2 + \delta_{\xi_2} \xi_1 
\label{modified Lie bracket}
\end{equation} that takes into account the possible field-dependence of the vector fields \eqref{residual 2d}, we obtain the commutation relations $[\xi(\epsilon_1, \chi_1, \eta_1),\xi(\epsilon_2, \chi_2, \eta_2)]_\star = \xi(\epsilon_{12}, \chi_{12}, \eta_{12})$, where
\begin{equation}
    \begin{split}
     \epsilon_{12}&=\epsilon_1\pu \epsilon_2 -\delta_{\xi_1} \epsilon_2 -(1\leftrightarrow 2), \\
      \eta_{12}&=\eta_1\chi_2 +\epsilon_1\pu \eta_2 -\delta_{\xi_1} \eta_2 -(1\leftrightarrow 2),\\
       \chi_{12}&=\epsilon_1\pu \chi_2 - \delta_{\xi_1} \chi_2 -(1\leftrightarrow 2)\,.
       \end{split} \label{commutation rel 2d}
\end{equation} Under residual gauge transformations generated by \eqref{residual 2d}, the solution space transforms infinitesimally as 
\begin{equation}
\begin{split}
    \delta_\xi A_0 &=\chi+\epsilon \pu A_0 +\pu \epsilon,  \qquad
    \delta_\xi Q_0 = -\chi +\epsilon \pu Q_0 ,\\
     \delta_\xi B_0 &= \epsilon \pu B_0+e^{-Q_0}(\eta\pu Q_0+e^{Q_0} B_0\, \pu \epsilon-\pu\eta),\\
     \delta_\xi \jf_0 &=e^{-Q_0}(\eta+e^{Q_0}\epsilon \pu \jf_0) , \qquad
     \delta_\xi \mathcal M =0  \,.
     \label{VARIATIONS 2D}
    \end{split}
\end{equation}
From general considerations,  \cite{Henneaux:1992ig , Barnich:2010xq , Compere:2018aar , Ruzziconi:2019pzd}, the action of the residual gauge diffeomorphisms on the solution space satisfies
\begin{equation}
    [\delta_{\xi_1} ,\delta_{\xi_2}] \alpha = - \delta_{[\xi_1, \xi_2]_\star} \alpha
    \label{Lie algebroid}
\end{equation} where $\alpha = (A_0, Q_0, B_0, \varphi_0)$, $ [\delta_{\xi_1} ,\delta_{\xi_2}] = \delta_{\xi_1} \delta_{\xi_2} - \delta_{\xi_2} \delta_{\xi_1}$ and $[\xi_1,\xi_2]_\star $ is the modified Lie bracket \eqref{modified Lie bracket} yielding \eqref{commutation rel 2d}. This equation highlights the structure of Lie algebroid \cite{Crainic} involved in the asymptotic symmetries \cite{Barnich:2010xq , Barnich:2017ubf}, where the base space is the solution space parametrized by $\alpha= (A_0, Q_0, B_0, \varphi_0)$, the Lie algebra at each point is formed by the span of the generators of the residual gauge diffeomorphisms $\xi(\epsilon, \chi, \eta )$ given in \eqref{residual 2d} and endowed with the bracket \eqref{modified Lie bracket}. In this picture, the relation \eqref{Lie algebroid} translates the fact that the anchor map $\xi \to \delta_\xi$ preserves the bracket between the Lie algebra and the tangent space at each point. 

\subsubsection{Dirichlet boundary conditions}
The general framework that we have presented here encompasses all the solution space analyses performed in the Eddington-Finkelstein types of gauge (see \textit{e.g} \cite{Afshar:2019axx ,Godet:2020xpk} and references therein). The more restrictive Dirichlet boundary conditions are imposed in linear dilaton Bondi gauge by requiring 
\begin{equation}
       A_0 =0, \qquad  Q_0 = 0\,.
    \label{Dirichlet 2d}
\end{equation} The residual gauge diffeomorphisms \eqref{residual 2d} preserving these boundary conditions are those whose parameters satisfy the constraints
\begin{equation}
     \partial_u \epsilon = 0, \qquad \chi = 0 
     \label{consequence Dirichlet}
\end{equation} In this case, the asymptotic symmetry algebra is $\mathbb{R}^{(\epsilon)} \oplus   C^{(\eta)}_\infty(\mathbb{R})$.




\subsection{Renormalization of the phase space}
In this section, we renormalize the action and the symplectic structure for the generic model \eqref{DGT} with vanishing kinematic potential in linear dilaton Bondi gauge. 
\subsubsection{Holographic renormalization}
\label{sec:Holographic renormalization 2d}

Before constructing the renormalized phase space of the theory, let us consider the minimal variational principle satisfying the following criteria (see the discussion in section \ref{sec:Conservation and variational principle}):
\begin{itemize}
    \item The action is finite on-shell, \textit{i.e.} $S= \mathcal{O}(r^0)$.
    \item The on-shell action has a well-defined flat limit in the sense discussed in section \ref{sec: Flat limit 2d}.
    \item When restricting our general framework to Dirichlet boundary conditions \eqref{Dirichlet 2d}, the action is stationary on solutions, \textit{i.e.} $\delta S = 0$.  
\end{itemize} In presence of timelike boundary (\textit{e.g.} for JT gravity \eqref{JT model}), one can show by an explicit computation that the following variational principle satisfies these requirements:
\begin{equation}
\begin{split}
    S =& \frac{1}{16\pi G} \int_M d^2 x \sqrt{-g}  [R X - 2V(X)]+ \int_{\partial M} du \, L_{GHY} \\
    &+   \int_{\partial M} du \, L_w + \int_{\partial M} du \, L_c + \int_{\partial M} du \, L_n
    \label{renormalized action 2d}
    \end{split}
\end{equation} where the first term is the bulk action for the dilaton gravity models \eqref{DGT}. The second term is the Gibbons-Hawking-York boundary term which is given explicitly by
\begin{equation}
     L_{GHY} =  \frac{1}{8\pi G}  \sqrt{|\gamma|}X K
     \label{GHY2D}
\end{equation} on each leaf of the foliation $r = \text{constant}$. Here $\gamma$ is the determinant of the metric $\gamma_{uu} du^2 = 2 B du^2$ induced on the leaves, and $K$ is the extrinsic curvature, $K = g^{\mu\nu} \nabla_\mu n_\nu$, with $n_\mu = \frac{1}{\sqrt{g^{rr}}} \delta_\mu^r=\frac{e^A}{\sqrt{|\gamma|}}\delta_\mu^r$ the unit normal vector. The second line in \eqref{renormalized action 2d} is given by
\begin{equation}
    L_w + L_c + L_n =\frac{1}{4\pi G} e^{A_0 + Q_0} \left(  \int^X dY V(Y)   \right) -\frac{1}{8\pi G} \frac{e^{A_0}}{\sqrt{|\gamma|}} \partial_u \left( X e^{-A_0} \sqrt{|\gamma|} \right) \, .
    \label{last line action}
\end{equation} These terms can be rewritten in a more covariant way as follows. We introduce the quantity 
    \begin{equation}
    w(X)=2\left(-\int^X dY V(Y)+e^{-(A_0 + Q_0)}\pu X \right) 
\end{equation}  in terms of  which the determinant of the induced boundary metric is
\begin{equation}
     \sqrt{|\gamma|}=e^{A_0+Q_0} \sqrt{\left |w\left(1-\frac{2\mathcal M}{w}\right)\right |} \,.
\end{equation} Assuming asymptotic dilaton domination \cite{Bagchi:2014ava}, $\frac{\mathcal{M}}{w(X)} \to 0$ when $r\to \infty$, which is the case for JT \eqref{JT model} and CGHS \eqref{CGHS model} models, the expression \eqref{last line action} can be rewritten as the sum of three covariant boundary Lagrangians. Indeed, we have explicitly 
\begin{equation}
\begin{split}
   &L_w = -\frac1{8\pi G}\sqrt{|\gamma|} \sqrt{|w|},  \qquad  L_c = \frac1{8\pi G} \sqrt{|\gamma|} D_a(v^a\, X), \\ &L_n= - \frac1{8\pi G} \sqrt{|\gamma|}X n_\mu v^b \partial_b n^\mu\,.
   \end{split}
   \label{covariant version}
\end{equation} Here, we write $(x^a) =(u)$ the unique coordinate on each leaf of the foliation $r = \text{constant}$, $v^a$ the unit vector field ($v^a v_a = \gamma_{uu} v^u v^u = -1$) tangent to the leaves and $D_a$ the covariant derivative with respect to the induced metric $\gamma_{uu} du^2$. Notice that the second term in \eqref{covariant version} is a corner Lagrangian \cite{Compere:2020lrt , Detournay:2014fva }. Indeed, we have $L_c = \frac1{8\pi G} \sqrt{|\gamma|} D_a(v^a\, X) = \frac1{8\pi G} \partial_a (\sqrt{|\gamma|} v^a\, X)$. 

The variational principle \eqref{renormalized action 2d} corresponds to the renormalized action proposed in \cite{Grumiller:2007ju} once considering the same regimes. In particuler, the Lagrangian $L_w$ in \eqref{covariant version} reduces to the counter-term proposed in \cite{Grumiller:2007ju} once requiring stationarity.  
Finally, notice that our action differs from the one proposed in \cite{Grumiller:2017qao} for JT gravity since we are considering open systems and assuming $\delta S = 0$ only in the Dirichlet case. 

When evaluated on-shell, the renormalized action \eqref{renormalized action 2d} is finite and reads as 
    \begin{equation}
    S=-\frac1{8\pi G}\int du \left(\jf_0\pu Q_0+\pu \jf_0 \right)-\Gamma_{bulk}(r_0)
\end{equation} where $\Gamma_{bulk}(r_0)$ is the finite contribution of the on-shell bulk action evaluated on its lower bound. The Euclidean version of this action can be interpreted as the free energy and used for thermodynamical considerations (see \textit{e.g.} \cite{Grumiller:2007ju}).

\subsubsection{Renormalization of the symplectic structure}

Inserting the on-shell Bondi metric \eqref{Bondi metric 2d} and the linear dilaton \eqref{Linear dilaton} into \eqref{presymplectic}, the radial component of the canonical presymplectic potential yields some divergences, which can be subtracted using the ambiguities of the covariant phase space formalism \eqref{ambiguity in theta}. We define the renormalized presymplectic potential \cite{Compere:2008us , Compere:2020lrt , Fiorucci:2020xto} as
\begin{equation}
    \Theta_{ren}^r [\phi; \delta \phi] = \Theta_{DGT}^r [\phi; \delta \phi] +\delta \left(L_{GHY}[\phi]+ L_w[\phi]+L_c[\phi]+L_n[\phi] \right)-\partial_a Y^{ar}  [\phi; \delta \phi]
    \label{renormalization potential 2d}
\end{equation}
where 
\begin{equation}
    Y^{ar}[\phi; \delta \phi] 
    = \frac{1}{2} r \bar{\Theta}^a_n[\phi; \delta \phi], \qquad \bar{\Theta}^a_n[\phi; \delta \phi] =  \frac{1}{8 \pi G} e^{-Q_0}\delta A_0\, .
    \label{ambig 2ddd}
\end{equation} As discussed in section \ref{sec:Conservation and variational principle}, the $\delta$-exact ambiguities in $ \Theta_{ren}^r [\phi; \delta \phi]$ are completely fixed by the boundary terms in the variational principle \eqref{renormalized action 2d}. Let us provide a covariant interpretation of the ambiguity \eqref{ambig 2ddd} in terms of the boundary structure. We define the boundary Lagrangian
\begin{equation}
    \bar{L}_n [A_0; \bar{\gamma}, \bar{X}]= - \frac1{8\pi G} \sqrt{|\bar{\gamma}|} \bar{X} \bar{n}_\mu \bar{v}^b \partial_b \bar{n}^\mu
    \label{boundary lag 2d}
\end{equation} as the pull-back of the Lagrangian $L_w$ defined in \eqref{covariant version} on the spacetime boundary ($\bar{X}$, $\bar{\gamma}$, $\bar{n}^\mu$, $\bar{n}_\mu$, $\bar{v}^a$ are the 
``unphysical" quantities associated with ${X}$, ${\gamma}$, ${n}^\mu$, ${n}_\mu$, ${v}^a$, respectively, through the conformal compactification process). The counter-term $\bar{\Theta}^a_n[\phi; \delta \phi]$ defined in \eqref{ambig 2ddd} corresponds to the presymplectic potential associated with \eqref{boundary lag 2d} where 
 the boundary metric $\bar{\gamma}_{uu}du^2$ is kept fixed 
 (\textit{i.e.} $\bar{\gamma}_{uu}du^2$ 
 is seen as background structure). 

 The explicit expression of the renormalized presymplectic potential \eqref{renormalization potential 2d} is finite and given by
\begin{equation}
    \Theta_{ren}^r[\phi; \delta \phi]  = -\frac{1}{8\pi G} \Big(  \delta B_0 + B_0 \delta (Q_0 + A_0) + \frac{1}{2} \delta A_0 \partial_u \varphi_0 - \frac{1}{2} \varphi_0 \partial_u \delta A_0 +\delta \varphi_0 \partial_u  Q_0 +\pu \delta \jf_0 \Big) 
\end{equation} from which one can derive the renormalized presymplectic current 
\begin{equation}
\begin{split}
    \omega^r_{ren} [\phi; \delta_1 \phi, \delta_2 \phi] =-\frac{1}{8\pi G} \Big[ \delta_2 B_0 \delta_1 (Q_0 + A_0) &+ \frac{1}{2} \delta_1 A_0 \partial_u \delta_2 \varphi_0  \\
    &-\delta_2 \varphi_0 \partial_u \delta_1 \Big(\frac{1}{2}  A_0 +   Q_0 \Big) \Big] 
    - (1 \leftrightarrow 2 ) \,.
    \end{split} \label{renormalized presymplectic current 2d}
\end{equation} An important observation is that the symplectic structure does not depend on the particular form of $V(X)$ in \eqref{DGT}. Therefore, the analysis of the charge algebra that we perform in the next section is very general and is not sensitive to the particular model of dilaton gravity that one is considering\footnote{However, as mentioned earlier, we have set the kinetic potential to zero in our analysis, \textit{i.e.} $U(X)= 0$. As can be seen from \eqref{BB codimension 2}, considering a non-vanishing kinetic potential could have an impact on the phase space analysis that would make it model-dependent. This justifies \textit{a posteriori} our choice of assumption.}. 

Finally, notice that \eqref{renormalized presymplectic current 2d} vanishes when we impose Dirichlet boundary conditions \eqref{Dirichlet 2d}. Hence, the associated charges are conserved and the variational principle \eqref{renormalized action 2d} is stationary on solutions, which is in agreement with our general discussion in section \ref{sec:Conservation and variational principle}.

\subsection{Integrability and charge algebra}
\label{sec:Integrability and charge algebra 2d}
In this section, we discuss the renormalized charges and present a particular slicing of the phase space for which they are integrable.  Furthermore we compute the charge algebra. 

\subsubsection{Surface charges}
\label{Surface charges in 2d}

The renormalization of the presymplectic potential \eqref{renormalization potential 2d} does not affect the finite part in $r$ of the Iyer-Wald charges. Furthermore, in the linear dilaton Bondi gauge (see equations \eqref{Bondi metric 2d} and \eqref{Linear dilaton}), the Barnich-Brandt and the canonical Iyer-Wald co-dimension $2$ forms coincide (\textit{i.e.} $E^{ru}[\phi; \delta_1\phi, \delta_2 \phi ] =0$, see \eqref{Link IW BB 2d} and \eqref{Link IW BB 2d v2}). Therefore, the finite expressions that we discuss now precisely correspond to the finite part of the Barnich-Brandt charges as well.

The renormalized co-dimension $2$ form can be derived from 
\begin{equation}
    \partial_u k^{ru}_{ren, \xi} [\phi; \delta \phi]= \omega^r_{ren} [\phi; \delta_\xi \phi, \delta \phi]
    \label{funda rel 2d}
\end{equation} (see \eqref{breaking nin conservation}), where $\omega^r_{ren} [\phi; \delta_1 \phi, \delta_2 \phi]$ is given explicitly in \eqref{renormalized presymplectic current 2d}. In two dimensions, a co-dimension $2$ surface is a point. The surface charges are therefore simply obtained by evaluating the co-dimension $2$ form at a point on the spacetime boundary
\begin{equation}
    \ndelta Q_\xi [\phi] = k^{ru}_{ren, \xi} [\phi; \delta \phi]\,.
\end{equation} The explicit expression reads as
\begin{align} \nonumber
    \ndelta Q_\xi [\phi] = \frac 1{16\pi G} \Big[  \delta \jf_0  (\chi-\partial_u \epsilon) +& e^{-Q_0}(\delta A_0+2 \delta Q_0)\eta  \\&+ 2\epsilon \Big(e^{Q_0+A_0}\delta \mathcal M +\frac12(\pu\jf_0\delta A_0- \pu A_0\delta \jf_0 )\Big)
\Big]\,. \label{charges non integrable 2d}
\end{align}
The charges are finite, thanks to the renormalization procedure \eqref{renormalization potential 2d}. As a consequence of \eqref{funda rel 2d}, they are generically not conserved.  Furthermore, the charges seem to be non-integrable. However, as explained in section \ref{sec:Integrability}, this apparent obstruction for integrability can be removed by performing field-dependent redefinitions of the symmetry parameters, which amounts to solve the Pfaff problem \cite{Barnich:2007bf , Adami:2020ugu ,  Compere:2017knf , Grumiller:2019fmp ,  Alessio:2020ioh, Ciambelli:2020shy}. In the present situation, we perform the redefinition
   \begin{equation}
      \begin{split}
        \eta&=e^{2Q_0+\frac12 A_0} \tilde \eta -\tilde \epsilon  \pu \jf_0 e^{-A_0}-\tilde \chi e^{-\frac12A_0} \jf_0 \, , \qquad
        \epsilon=   e^{-(Q_0+A_0)}\, \tilde \epsilon \, ,\\
        \chi & = 2 e^{-(Q_0+\frac12A_0)} \tilde \chi +e^{-(Q_0+A_0)}(  \pu Q_0\, \tilde \epsilon +\pu \tilde \epsilon)
        \label{new parameters}
        \end{split}
    \end{equation}
    where $\tilde{\epsilon}$, $\tilde{\eta}$ and $\tilde{\chi}$ are taken to be field-independent, \textit{i.e.} $\delta \tilde{\epsilon} = \delta\tilde{\eta} =\delta\tilde{\chi} = 0$. In terms of these parameters, the commutation relations \eqref{commutation rel 2d} become $[\xi(\tilde{\epsilon}_1, \tilde{\eta}_1 , \tilde{\chi}_1), \xi (\tilde{\epsilon}_2, \tilde{\eta}_2 , \tilde{\chi}_2)]_\star =  \xi(\tilde{\epsilon}_{12}, \tilde{\eta}_{12} , \tilde{\chi}_{12}) $ with 
    \begin{equation}
      \tilde{\epsilon}_{12} =  \tilde{\eta}_{12} = \tilde{\chi}_{12} = 0 \, .
      \label{symmetry algebra 2d}
    \end{equation} Henceforth, the algebra of the residual gauge diffeomorphisms is abelian and is given by $C_{\infty}^{(\epsilon)}(\mathbb{R})  \oplus C^{(\eta)}_{\infty}(\mathbb{R}) \oplus C_{\infty}^{(\chi)}(\mathbb{R})$. Moreover, the redefinition \eqref{new parameters} renders the charges \eqref{charges non integrable 2d} integrable, namely $\ndelta Q_\xi [\phi] \equiv \delta Q_\xi [\phi]$ with
\begin{align}
   \delta Q_\xi [\phi]&=\frac{1}{8\pi G}\Big[ \tilde \epsilon \, \delta\mathcal M +\tilde \chi\, \delta(e^{-(Q_0+\frac12A_0)} \jf_0)+\tilde \eta \,\delta( e^{Q_0+\frac12 A_0})\Big]\,.
\end{align} Integrating this expression on a path in the solution space gives the finite charge expression
\begin{equation}
     Q_\xi [\phi] =\frac{1}{8\pi G}\Big[ \tilde \epsilon \, \mathcal M +\tilde \chi\, (e^{-(Q_0+\frac12A_0)} \jf_0)+\tilde \eta \,( e^{Q_0+\frac12 A_0})\Big]\,.
     \label{final charge 2d}
\end{equation} 
The charges involve three independent combinations of the solution space, one of them being the particular function that corresponds to the Casimir \eqref{2dCasimir}. In terms of the new parameters \eqref{new parameters}, the variations of the combinations of the solution space appearing in the charges \eqref{final charge 2d} are extremely simple
\begin{equation}
        \delta_\xi \mathcal{M} =0 , \qquad \delta_\xi (e^{\frac{1}{2} A_0 + Q_0}) = - \tilde{\chi}, \qquad
        \delta_\xi ( e^{-\frac{1}{2} A_0 - Q_0} \varphi_0) = \tilde{\eta}\,. 
        \label{simple variations 2d}
\end{equation} 
The integrable slicing \eqref{new parameters} of the phase space renders manifest the split between the directions tangent to the orbits $\mathcal M= \text{constant}$ and the transverse direction related to the exact symmetry $\xi = \tilde \epsilon$ with $\partial_u \tilde \epsilon = 0$ \cite{Compere:2015bca}. 

Notice that among all the possible field-dependent redefinitions of the parameters that render the charges integrable, the choice \eqref{new parameters} has the property that we can impose the Dirichlet boundary conditions \eqref{Dirichlet 2d} consistently with the requirements to keep the parameters field-independent. This can be readily seen from the variation of $A_0$ and $Q_0$ in terms the new parameters:
\begin{equation}
    \delta_\xi A_0= 2e^{-(Q_0+A_0)} \left( e^{\frac12 A_0}\tilde \chi +\pu \tilde \epsilon\right), \qquad
\delta_\xi Q_0= e^{-(Q_0+A_0)} \left(2 e^{\frac12 A_0}\tilde \chi +\pu \tilde \epsilon\right)\,.
\end{equation} Indeed, preserving the Dirichlet boundary conditions \eqref{Dirichlet 2d} implies $\partial_u \tilde{\epsilon} = 0$ and $\tilde{\chi}=0$. Notice that in this case, the charges are simply $Q_\xi[\phi]=\frac1{8\pi G} \tilde \epsilon \, \mM$. In particular they are conserved 
as discussed below \eqref{renormalized presymplectic current 2d}.

\subsubsection{Charge algebra}

The charges \eqref{final charge 2d} being integrable, one can use the representation theorem which states that the charges form a representation of the asymptotic symmetry algebra, up to a possible central extension \cite{Brown:1986nw , Brown:1986ed} (see also \cite{Barnich:1991tc, Barnich:2001jy, Barnich:2007bf} for the covariant formulation of this result). Indeed, using the Peierls bracket \cite{Peierls:1952cb , Harlow:2019yfa} 
\begin{equation}
    \{ Q_{\xi_1}[\phi] , Q_{\xi_2}[\phi] \} \equiv \delta_{\xi_2} Q_{\xi_1}[\phi]
\end{equation}  and the variations \eqref{simple variations 2d}, one can show that the charges \eqref{final charge 2d} satisfy
\begin{equation}
    \{ Q_{\xi_1}[\phi] , Q_{\xi_2}[\phi] \} = \frac{1}{8\pi G} (\tilde{\chi}_1 \tilde{\eta}_2 - \tilde \chi_2 \tilde \eta_1 )\,.
    \label{charge algebra 2d}
\end{equation} The central extension appearing in the right-hand side of this charge algebra is a non-trivial. Indeed, the $2$-cocycle of an abelian algebra cannot be a coboundary. Therefore, the charge algebra is given by the direct sum $C_\infty (\mathbb{R}) \oplus \text{Heisenberg}$, \textit{i.e.} the charges represent the algebra of symmetries $C_{\infty}^{(\epsilon)}(\mathbb{R})  \oplus C^{(\eta)}_{\infty}(\mathbb{R}) \oplus C_{\infty}^{(\chi)}(\mathbb{R})$ displayed in \eqref{symmetry algebra 2d}, up to a central extension that appears in the $C^{(\eta)}_{\infty}(\mathbb{R}) \oplus C_{\infty}^{(\chi)}(\mathbb{R})$ sector. 

These general results echo some previous analyses in lower-dimensional gravity concerning the asymptotic structure of the spacetime near generic null hypersurfaces \cite{Adami:2020ugu}. It was found in that reference that the charges in two dimensions can always been made integrable by a field-dependent redefinition of the parameters. Furthermore, it was shown that the Heisenberg algebra always appears when considering the most general boundary conditions around null hypersurfaces in the bulk. Therefore, our work confirms and extends these results to timelike and null hypersurfaces at infinity. Furthermore, it exhibits the presence of the Casimir labelling the orbits in the phase space. 

Notice that the signature of the spacetime boundary will depend on the particular model that we consider. In JT gravity \eqref{JT model}, the boundary is timelike, while in CGHS gravity \eqref{CGHS model}, the boundary is null. The interplay between these two models is discussed in the next section. 

Moreover, notice that the partial gauge fixing \eqref{Bondi metric 2d} which is reached by using only one out of the two diffeomorphism degrees of freedom in two dimensions allows us to find additional asymptotic symmetries. Indeed, consider the additional condition 
\begin{equation}
    A_0 = 0 \quad \Longleftrightarrow \quad g_{ru} = -1
    \label{NU gauge}
\end{equation} which fixes the gauge completely for the metric \eqref{Bondi metric 2d}. In this case, the linear dilaton condition \eqref{Linear dilaton} is just a consequence of the equations of motion and does not impose further restriction, see \eqref{coordradial}. The additional gauge-fixing condition \eqref{NU gauge} is preserved under residual gauge diffeomorphisms if the parameters satisfy
\begin{equation}
    \tilde \chi = - \partial_u \tilde\epsilon\,.
\end{equation} Henceforth, we loose one of the three independent symmetry parameters. Consequently, we loose one tower of charges since \eqref{final charge 2d} becomes  
\begin{equation}
     Q_\xi [\phi] =\frac{1}{8\pi G}\Big[ \tilde \epsilon \, \mathcal M - \partial_u \tilde\epsilon \, (e^{-(Q_0+\frac12A_0)} \jf_0)+\tilde \eta \,( e^{Q_0+\frac12 A_0})\Big]
\end{equation} and the asymptotic symmetry algebra reduces to $C_\infty^{(\epsilon)}(\mathbb{R}) \oplus C_\infty^{(\eta)}(\mathbb{R})$. This observation that gauge fixing eliminates potentially interesting asymptotic symmetries confirms previous analyses performed in different contexts  \cite{Adami:2020amw ,Grumiller:2016pqb , Grumiller:2017sjh , Ciambelli:2020eba , Ciambelli:2020ftk}. The charge algebra \eqref{charge algebra 2d} reduces to the Heinsenberg algebra, while the central charge reads as 
\begin{equation}
    \frac{1}{8\pi G} (\partial_u \epsilon_1 \eta_2 - \partial_u \epsilon_2 \eta_1)\,.
\end{equation} Furthermore, when restricted to JT gravity, our analysis encompasses the recent results concerning new boundary conditions in asymptotically AdS$_2$ spacetimes \cite{Godet:2020xpk}. In particular, our study provides the field-dependent redefinition of the parameters that renders the charges discussed in that reference integrable. 


\subsection{Flat limit}
\label{sec: Flat limit 2d}

The JT dilaton gravity model \eqref{JT model} is the analogue of the higher-dimensional gravity theories with non-vanishing cosmological constant. Indeed, it contains asymptotically (A)dS$_2$ spacetime solutions, including black holes.  The solution space discussed in section \ref{Linear Dilaton Bondi gauge in two dimensions} specified for JT gravity \eqref{JT model} reads as
\begin{equation}
    ds^2 = 2 B(u,r) du^2 - 2 e^{A_0(u)} du dr, \qquad X = e^{-Q_0(u)} r + \varphi_0(u)
    \label{JT1}
\end{equation} where 
\begin{equation}
    B(u,r) = \frac{r^2}{2} \Lambda e^{2 A_0} + r e^{A_0} ( \Lambda e^{A_0+ Q_0 } \varphi_0 +  \partial_u Q_0 ) + \frac{\Lambda}{2} \varphi_0^2 e^{2(A_0 + Q_0)} + B_0 e^{(Q_0 + A_0)}\,.
    \label{JT2}
\end{equation}

However, in contrast with gravity in higher dimension, taking naively the limit $\Lambda \to 0$ in JT gravity \eqref{JT model} does not yield an interesting model. In fact, the solution space of the model
\begin{equation}
    L[\phi] = \frac{\sqrt{-g}}{16\pi G} RX
\end{equation} is given by  
\begin{equation}
    ds^2 = 2  e^{A_0} (r  \partial_u Q_0  + B_0 e^{Q_0 }) du^2 - 2 e^{A_0(u)} du dr, \qquad X = e^{-Q_0(u)} r + \varphi_0(u)\,.
    \label{globally Minkowski}
\end{equation} Despite it contains two-dimensional Minkowski space as a solution ($Q_0 = 0 = A_0$, $B_0 = -\frac{1}{2}$), it does not contain any black hole solution.  Therefore, we consider another road to obtain models with interesting asymptotically flat solutions. Starting from \eqref{JT model} and shifting the dilaton by a constant $X \to X + \alpha$, the Lagrangian reads as
\begin{equation}
    L_{JT}[\phi] = \frac{\sqrt{-g}}{16\pi G} [(R +2 \Lambda) (X + \alpha)] = \frac{\sqrt{-g}}{16\pi G} [RX + 2 \Lambda \alpha + R \alpha + 2 \Lambda X  ]\,.
\end{equation} The third term in the last expression is the Gauss-Bonnet topological term in two dimensions and can be discarded since it will not contribute to the dynamics. Now, defining $\lambda = \alpha \Lambda$ and taking the limit
\begin{equation}
    \alpha \to \infty , \qquad  \Lambda \to 0, \qquad \lambda \text{ kept fixed},
    \label{flat limit 2d}
\end{equation} we precisely recover the CGHS model \eqref{CGHS model}, which contains interesting class of asymptotically flat black hole solutions \cite{Callan:1992rs}. Therefore, we will consider the limit \eqref{flat limit 2d} as the flat limit in two dimensions, relating asymptotically AdS$_2$ solutions in JT gravity to asymptotically flat solutions in CGHS gravity. The solution space discussed in section \ref{Linear Dilaton Bondi gauge in two dimensions} specified for CGHS gravity \eqref{CGHS model} reads as 
\begin{equation}
    ds^2 = 2 B(u,r) du^2 - 2 e^{A_0(u)} du dr, \qquad X = e^{-Q_0(u)} r + \varphi_0(u)
\end{equation} where 
\begin{equation}
    B(u,r) = r e^{A_0} ( \lambda e^{A_0+ Q_0 } +  \partial_u Q_0 ) +\lambda \varphi_0 e^{2(A_0 + Q_0)} + B_0 e^{(Q_0 + A_0)}\,.
\end{equation} This solution space can be obtained in the flat limit \eqref{flat limit 2d} of the solution space of JT gravity given in \eqref{JT1}-\eqref{JT2}, after re-absobring the $\alpha$ parameter in the dilaton. 

As discussed above, under the assumption $U(X) = 0$, the symplectic structure of the theory does not depend on the particular expression of the potential $V(X)$. Therefore, the analyses performed in section \ref{sec:Integrability and charge algebra 2d} is valid for both JT and CGHS gravity theories. Consequently, the flat limit of the phase space is immediate. Notice however that the interpretation of the asymptotic region for the two models is drastically different. In JT gravity, the boundary is timelike and is endowed with a boundary metric $\Lambda e^{2 A_0} du^2$, while in CGHS gravity, the boundary is null and endowed with a degenerate (vanishing) metric.

\section{Gravity in three dimensions}
\label{sec:Gravity in three dimensions}

We now apply the general framework presented in section \ref{Integrability and conservation in the covariant phase space formalism} to the case of three-dimensional Einstein gravity.

\subsection{Phase space of three-dimensional gravity}
\label{Phase space of three-dimensional gravity}

The phase space of general relativity in three dimensions has been extensively studied in the literature (see \textit{e.g.} \cite{Compere:2018aar, Compere:2007az, Ruzziconi:2019pzd, Riegler:2017fqv} for reviews). Let us briefly sketch the main results. We start from the Einstein-Hilbert Lagrangian density
\begin{equation}
    L_{EH}[g] = \frac{\sqrt{-g}}{16 \pi G} \left(R + \frac{2}{\ell^2} \right)\,.
\end{equation} Taking an infinitesimal variation of the Lagrangian yields 
\begin{equation}
    \delta L_{EH}[g] = \frac{\delta L_{EH}}{\delta g_{\mu\nu}} \delta g_{\mu \nu} + \partial_\mu \Theta^\mu_{EH} [g; \delta g]\,.
    \label{EH Theory}
\end{equation} The Euler-Lagrange derivatives
\begin{equation}
    \frac{\delta L_{EH}}{\delta g_{\mu\nu}} = -\frac{\sqrt{-g}}{16\pi G}\left( G^{\mu\nu} - \frac{1}{\ell^2} g^{\mu\nu}\right)
\end{equation} lead to the Einstein equations and the canonical Einstein-Hilbert presymplectic potential reads as
\begin{equation}
    \Theta^\mu_{EH} [g; \delta g] = \frac{\sqrt{-g}}{16\pi G} [\nabla_\nu (\delta g)^{\mu\nu} - \nabla^\mu (\delta g)^\nu_\nu]\,.
    \label{canonical}
\end{equation}

The Einstein-Hilbert theory \eqref{EH Theory} is invariant under diffeomorphisms which act on the metric with a standard Lie derivative $\delta_\xi g_{\mu\nu} = 2 \nabla_{(\mu} \xi_{\nu)}$. We have 
\begin{equation}
    \frac{\delta L_{EH}}{\delta g_{\mu\nu}} \delta_\xi g_{\mu \nu} =  \frac{\sqrt{-g}}{8\pi G} \nabla_\mu  G^{\mu\nu} \xi_\nu  + \partial_\mu S^\mu_\xi[g]
\end{equation} after integrating by parts to isolate the diffeomorphism parameters. The first term in the right-hand side vanishes because of the Bianchi identities which correspond to the Noether identities of the theory. The total derivative term contains the weakly-vanishing Noether current
\begin{equation}
    S^\mu_\xi [g] = 2 \frac{\delta L_{EH}}{\delta g_{\mu\nu}} \xi_\nu = - \frac{\sqrt{-g}}{8 \pi G} \left( G^{\mu\nu} - \frac{1}{\ell^2} g^{\mu\nu}\right) \xi_\nu\,.
\end{equation} Applying the homotopy operator \eqref{homotopy operator} on this expression gives the Barnich-Brandt co-dimension $2$ form (which is a $1$-form in three dimensions)
\begin{equation}
\begin{split}
    k^{\mu\nu}_{BB,\xi} [g; \delta g] = \frac{\sqrt{-g}}{8\pi G} \Big(& \xi^\mu \nabla_\sigma (\delta g)^{\nu\sigma} - \xi^\mu \nabla^\nu (\delta g)^\sigma_\sigma + \xi_\sigma \nabla^\nu (\delta g)^{\mu\sigma} \\
    &+ \frac{1}{2} (\delta g)^\sigma_\sigma \nabla^\nu \xi^\mu - \frac{1}{2} (\delta g)^{\sigma \nu} \nabla_\sigma \xi^\mu + \frac{1}{2} (\delta g)^\nu_\sigma \nabla^\mu \xi^\sigma  \Big)\,.
    \end{split}
    \label{BB charges in 3d}
\end{equation} It is defined up to an exact $1$-form $k^{\mu\nu}_{BB,\xi}[g;\delta g] \to k^{\mu\nu}_{BB,\xi}[g; \delta g] + \partial_\rho M_\xi^{[\mu\nu\rho]}[g; \delta g]$ that will play no role when integrating on compact co-dimension $2$ surface to obtain the gravitational charges.

The Barnich-Brandt and Iyer-Wald co-dimension $2$-forms are related through \eqref{Link IW BB 2d v2} where 
\begin{equation}
    E^{\mu\nu}[g; \delta_1 g, \delta_2 g] = \frac{\sqrt{-g}}{32\pi G} (\delta_1 g)^\mu_\sigma  (\delta_2 g)^{\sigma \nu} - (1 \leftrightarrow 2 )\,.
    \label{E in 3d}
\end{equation}
We have the conservation law 
\begin{equation}
    \partial_\nu k^{\mu\nu}_{\xi} [g; \delta g] = \omega^\mu_{EH}[g;\delta_\xi g , \delta g]
\end{equation} where $\omega^\mu_{EH}[g; \delta_1 g, \delta_2 g] = \delta_2 \Theta_{EH}^\mu  [g; \delta_1 g] - \delta_1 \Theta_{EH}^\mu[g; \delta_2 g] $. As in the two-dimensional case, we will use the Iyer-Wald procedure to derive the charges and play with the ambiguities \eqref{ambiguity in theta} to eliminate the divergences.

\subsection{Bondi gauge in three dimensions}

The Bondi gauge in three dimensions has been studied in \textit{e.g.} \cite{Barnich:2006av, Barnich:2010eb} to investigate asymptotically AdS$_3$ and asymptotically flat spacetimes. The analysis was then extended to asymptotically locally AdS$_3$ and asymptotically locally flat spacetimes to include the boundary structure in the solution space \cite{Ciambelli:2020eba, Ciambelli:2020ftk}. We review these results here. 

\subsubsection{Solution space}
Writing the coordinates as $(x^\mu) = (u,r,\phi)$, the Bondi gauge is obtained by requiring the following three gauge-fixing conditions 
\begin{equation}
g_{rr} = 0, \quad g_{r \phi} = 0, \quad g_{\phi \phi} = r^2  e^{2 \varphi} .
\label{Bondi gauge fixing 3d}
\end{equation} where $\varphi$ is function of $(u,\phi)$. This gauge is always reachable by using the three degrees of freedom we have on coordinate transformations. The Bondi gauge line element in three dimensions takes the form
\begin{equation}
d s^2 = \frac{V}{r} e^{2\beta} d u^2 - 2 e^{2 \beta} d u d r + r^2 e^{2 \varphi} (d \phi - U d u)^2 \,,
\label{Bondi metric}
\end{equation} In this expression, $V$, $\beta$ and $U$ are functions of $(u, r, \phi)$. Solving Einstein's equations $G_{\mu\nu} - \frac{1}{\ell^2} g_{\mu\nu} = 0$ gives the following expansions
\begin{equation}
    \begin{split}
\beta &= \beta_0 (u, \phi), \\
U &= U_0(u, \phi) + \frac{1}{r}  2 e^{2\beta_0} e^{-2 \varphi} \partial_\phi \beta_0  - \frac{1}{r^2} e^{2\beta_0} e^{-2 \varphi} N(u, \phi), \\
\frac{V}{r} &= -\frac{r^2}{\ell^2} e^{2 \beta_0} - 2 r (\partial_u \varphi + \partial_\phi U_0 + \partial_\phi \varphi U_0) + M (u, \phi) + \frac{1}{r} 4 e^{2 \beta_0} e^{-2 \varphi} N \partial_\phi \beta_0 - \frac{1}{r^2} e^{2 \beta_0} e^{-2 \varphi} N^2. \label{du N}
\end{split}
\end{equation} In these expressions, $M = M(u, \phi)$ is the Bondi mass aspect and $N=N(u,\phi)$ is the angular momentum aspect. The Einstein equations also lead to time evolution constraints on $M$ and $N$ (see \cite{Ciambelli:2020eba, Ciambelli:2020ftk} for the explicit expressions). 
The solution space is thus paramatrized by five arbitrary functions of $(u, \phi)$ with two time evolution constraints. Three of them ($\beta_0$, $U_0$, $\varphi$) characterize the induced boundary metric on the boundary $\mathscr{I}$ through
\begin{equation}
     \bar{\gamma}_{ab} dx^a dx^b \equiv \lim_{r\to \infty} \left( \frac{1}{r^2} ds^2 \right) = \left(-\frac{e^{4\beta_0}}{\ell^2} + e^{2\varphi} U_0^2\right) du^2 - 2 e^{2\varphi} U_0 du d\phi
    + e^{2\varphi} d\phi^2
    \label{induced boundary metric}
\end{equation} where $(x^a) = (u, \phi)$ are the coordinates on $\mathscr{I}$. The two other functions ($M$, $N$) encode the bulk information on the mass and the angular momentum.

\subsubsection{Residual symmetries}

The on-shell residual gauge diffeomorphisms preserving the Bondi gauge-fixing conditions \eqref{Bondi gauge fixing 3d} are generated by vector fields $\xi = \xi^u \partial_u + \xi^\phi \partial_\phi + \xi^r \partial_r$ whose components read explicitly as
\begin{equation}
\begin{split}
\xi^u =& f,\\
\xi^\phi =& Y - \frac{1}{r} \partial_\phi f\, e^{2\beta_0-2\varphi} ,\\
\xi^r =& - r [ \partial_\phi Y - \omega - U_0 \partial_\phi f + Y \partial_\phi \varphi + f \partial_u \varphi ] \\
&+ e^{2\beta_0 - 2 \varphi} (\partial_\phi^2 f - \partial_\phi f \partial_\phi \varphi + 4 \partial_\phi f \partial_\phi \beta_0) - \frac{1}{r} e^{2\beta_0 - 2 \varphi}  \partial_\phi f\, N \label{resBON3}
\end{split}
\end{equation} where $f$, $Y$ and $\omega$ are arbitrary functions of $(u,\phi)$ that may be field-dependent \cite{Ciambelli:2020eba, Ciambelli:2020ftk}. Using the modified Lie bracket \eqref{modified Lie bracket}, these vector fields satisfy the commutation relations $[\xi(f_1, Y_1, \omega_1), \xi(f_2, Y_2, \omega_2)]_\star = {\xi}({f}_{12}, {Y}_{12},{\omega}_{12})$ where
\begin{equation}
    \begin{split}
       & {f}_{12} = f_1 \partial_u f_2 + Y \partial_\phi f_2 - \delta_{\xi_1} f_2 - (1 \leftrightarrow 2) \, ,\\
        & {Y}_{12} = f_1 \partial_u Y_2 + Y_1 \partial_\phi Y_2 - \delta_{\xi_1} Y_2 - (1 \leftrightarrow 2) \, ,\\
        & {\omega}_{12} = - \delta_{\xi_1} \omega_2 - (1 \leftrightarrow 2) \, .  \\
    \end{split} \label{commutation relations 3d}
\end{equation} The terms $\delta_\xi f$, $\delta_\xi Y$ and $\delta_\xi \omega$ are present to take into account the possible field-dependence of the parameters. Indeed, as explained below, a field-dependent redefinition of these parameters will be necessary to make the charges integrable.

Under these infinitesimal residual gauge diffeomorphisms, the boundary structure transforms as
\begin{equation}
    \begin{split}
\delta_\xi \varphi =& \omega, \\
\delta_\xi \beta_0 =& (f \partial_u + Y \partial_\phi)\beta_0 + \left(\frac{1}{2}\partial_u - \frac{1}{2} \partial_u \varphi + U_0 \partial_\phi \right) f - \frac{1}{2}(\partial_\phi Y + Y \partial_\phi \varphi - \omega ), \\
\delta_\xi U_0 =& (f \partial_u + Y \partial_\phi - \partial_\phi Y ) U_0 - \left(\partial_u Y - \frac{1}{\ell^2} e^{4 \beta_0} e^{-2 \varphi} \partial_\phi f\right) + U_0 (\partial_u f +  U_0 \partial_\phi f) . \label{variation of the boundary structure 3d}
\end{split}
\end{equation} The explicit variations of the angular momentum and mass aspects can be found in \cite{Ciambelli:2020eba, Ciambelli:2020ftk}. 

\subsubsection{Dirichlet boundary conditions}

Let us mention that this general framework contains all the previous analyses performed in Bondi gauge. In particular, Dirichlet boundary conditions \cite{Brown:1986nw} are imposed by requiring 
\begin{equation}
    \beta_0 = 0, \qquad U_0 = 0, \qquad \varphi = 0
    \label{Dirichlet 3d}
\end{equation} on the solution space \cite{Barnich:2012aw}. This implies that the induced boundary metric \eqref{induced boundary metric} is flat. From \eqref{variation of the boundary structure 3d}, the residual gauge diffeomorphisms \eqref{resBON3} preserving \eqref{Dirichlet 3d} are those whose parameters satisfy the constraints
\begin{equation}
    \partial_u f = \partial_\phi Y , \qquad \partial_u Y = \frac{1}{\ell^2} \partial_\phi f,\qquad \omega = 0\,, \label{ckv 3d}
\end{equation} \textit{i.e.} they induce conformal Killing vectors on the boundary.

\subsection{Renormalization of the phase space}

Similarly to what happens in the two-dimensional case, when evaluating the radial component of the canonical Einstein-Hilbert presymplectic potential \eqref{canonical} on the solution space displayed in the previous section, we obtain some $\mathcal{O}(r^2)$ terms that diverge when $r\to \infty$. Furthermore, the presymplectic potential also admits some $\mathcal{O}(\ell^2)$ terms that constitute an obstruction to take the flat limit $\ell \to \infty$. Therefore, to eliminate these divergences, we have to add some counter-terms using the ambiguities \eqref{ambiguity in theta} of the formalism. We show below that the precise form of these divergences are related to the counter-terms that one has to add to the action in the holographic renormalization process \cite{deHaro:2000vlm , Compere:2008us}. Moreover, we show that the counter-terms do not modify the finite part of the charges.

\subsubsection{Holographic renormalization in Bondi gauge}
\label{Holographic renormalization in Bondi gauge}

As discussed in section \ref{sec:Conservation and variational principle}, a variational principle $S$ will be appropriate if it satisfies the three following requirements:
\begin{itemize}
\item The action is finite on-shell, \textit{i.e.} $S = \mathcal{O}(r^0)$.
\item The on-shell action has a well-defined flat limit, \textit{i.e.} $S= \mathcal{O}(\ell^0)$
    \item When restricting our general framework to Dirichlet boundary conditions \eqref{Dirichlet 3d}, the action is stationary on solutions, \textit{i.e.} $\delta S = 0$.
\end{itemize} In asymptoticaly locally AdS$_3$ spacetimes, the minimal action principle in Bondi gauge that satisfies these requirements is given by
\begin{align}
    S=& \frac1{16\pi G}\int_{ M}\sqrt{-g} \left(R+ \frac{2}{\ell^2} \right) d^3x  + \int_{\partial M} a_1 L_{GHY} ~d^2 x + \int_{\partial M} a_2 L_{ct} ~d^2 x \nonumber\\& + \int_{\partial M} a_3 \, 
    L_\circ ~d^2 x + \int_{\partial M} a_4 \, 
    L_b ~d^2 x  + \int_{\partial M} a_5  \, 
    L_R ~d^2 x
    \label{holographically renormalized action}
\end{align}
where the coefficients are settled to
\begin{equation}
    a_i =1
     \label{values of the coefficients}
\end{equation} with $i=1, \ldots , 5$. Let us now describe each term in this action and justifies that this is actually the right choice.

The first line of \eqref{holographically renormalized action} is the holographically renormalized variationnal principle that one would expect to have in Fefferman-Graham gauge \cite{deHaro:2000vlm}. The first piece is the Einstein-Hilbert buk action. The second piece is made of the Gibbons-Hawking-York boundary term
\begin{equation}
    L_{GHY} = \frac{1}{8 \pi G} \sqrt{-\gamma} K
\end{equation}
constructed from the extrinsic curvature $K = g^{\mu\nu} \nabla_{(\mu} n_{\nu)}$, where $n_\mu = \frac{1}{\sqrt{g^{rr}}} \delta_\mu^r$ is the unit normal vector to the foliation $\{r= \text{constant}\}$ induced by the Bondi gauge \eqref{Bondi gauge fixing 3d}\footnote{The leading terms of the extrinsic curvature are given by $
    K= \frac2\ell +\frac\ell {2 } R[\gamma] +\mathcal O(r^{-3}) $,
where $R[\gamma]$ is the Ricci scalar of the induced metric $\gamma_{ab}$.}. The coordinates on each leaf of the foliation are $(x^a) = (u,\phi)$. The induced metric is written $\gamma_{ab}$ and is obtained by taking the pull-back of
\begin{equation}
    \gamma_{\mu\nu} = g_{\mu\nu} - n_\mu n_\nu
\end{equation} on each leaf. The determinant and the Levi-Civita connection associated with the induced metric are written $\gamma = \det (\gamma_{ab})$ and $D_a$, respectively. The last term in the first line is the counter-term prescribed by the holographic renormalization in Fefferman-Graham gauge
\begin{equation}
    L_{ct} = - \frac{1}{8 \pi G \ell} \sqrt{-\gamma} \,.
\end{equation}

 These terms are not sufficient in Bondi gauge to remove the $r-$divergences and one has to add additional contributions displayed in the second line of \eqref{holographically renormalized action} that we describe now. The first term 
\begin{equation}
    L_\circ = -\frac{1}{8 \pi G} \sqrt{-\gamma} D_a v^a , \qquad v^a \partial_a = \frac{r \, e^\varphi}{\sqrt{-\gamma}} (\partial_u + U \partial_\phi)
    \label{corner lagrangian}
\end{equation} is a corner Lagrangian \cite{Detournay:2014fva , Compere:2020lrt}. Indeed, we have $L_\circ = \partial_a L^a_C$, where $L_C = -\frac{1}{8 \pi G}\sqrt{-\gamma} v^a$. The vector field $v^a$ appearing in \eqref{corner lagrangian} is tangent to the leaves of the foliation and satisfies the two properties
 \begin{equation}
    v^a \gamma_{ab} v^b = -1 , \qquad \lim_{r\to \infty} \left( \frac{1}{r} \gamma_{ab} v^b \right) = - \frac{e^{2 \beta_0}}{\ell} \delta^u_a \,.
   \label{properties va}
\end{equation} Its geometric interpretation is clear: this is the unique future-oriented unit vector that indicates the direction in which the boundary metric $\frac{1}{r^2} ds^2|_{\mathscr{I}}$ degenerates in the flat limit $\ell \to \infty$. The next term in \eqref{holographically renormalized action} is a kinetic term for this vector,
\begin{equation}
    L_b = \frac{\ell}{16 \pi G} \sqrt{-\gamma} (D_a v^a)^2\,.
\end{equation} Finally, the last term in \eqref{holographically renormalized action} is the Gauss-Bonnet term
\begin{equation}
    L_R = -\frac{\ell}{16 \pi G} \sqrt{-\gamma} R[\gamma]\,.
\end{equation} 


Let us now provide some details on how to obtain the values of the coefficients \eqref{values of the coefficients} of the various terms in the action \eqref{holographically renormalized action}. 
\begin{itemize}
    \item When evaluating the action \eqref{holographically renormalized action} on-shell,  divergences in $\mathcal{O}(r^2)$ arise. The $r^2$-divergences are removed by imposing $2 a_1 - a_2 = 1$, while the $r$-divergences are suppressed by $2 a_1 - a_2 - a_3 = 0$.
    \item When evaluated on-shell, the action \eqref{holographically renormalized action} exhibits some terms in $\mathcal{O}(\ell)$ that are eliminated by imposing $a_5 =  a_1$ and $2 a_1 - a_2 -a_4 =0$.
     \item The condition that the action is stationnary on solutions when Dirichlet boundary conditions \eqref{Dirichlet 3d} are imposed requieres $a_1 = 1$.
\end{itemize}
Putting all these constraints together, we obtain \eqref{values of the coefficients}. Sending the cut-off to infinity, $r \to \infty$, the expression of the on-shell renormalized action \eqref{holographically renormalized action} is explicitly given by 
\begin{equation}
    S =
    \frac1{16\pi G} \int d^2x \left( -e^{\varphi} M -4 e^{2\beta_0-\varphi} (2(\partial_\phi\beta_0)^2-\partial_\phi \beta_0\partial_\phi \varphi +\partial_\phi^2\beta_0)\right) -\Gamma_{bulk}(r_0)
    \label{on-shell action}
\end{equation}  where $\Gamma_{bulk}(r_0)$ is the finite contribution of the on-shell bulk action evaluated on its lower bound. One can check that the Euclidean version of the on-shell action \eqref{on-shell action}, when evaluated for BTZ black hole, exactly reproduces the Gibbs free energy obtained by Legendre transformation of the BTZ black hole mass (see \textit{e.g.}  \cite{Kraus:2006wn}). A similar computation can be done for flat space cosmologies \cite{Barnich:2012xq,Bagchi:2012xr} when considering the Euclidean version of the flat limit of \eqref{on-shell action}.

\subsubsection{Renormalization of the symplectic structure}

We now renormalize the presymplectic potential using the ambiguities of the covariant phase space formalism \eqref{ambiguity in theta}. The counter-terms to remove the $\mathcal{O}(r^2)$ and $\mathcal{O}(\ell^2)$ divergences are similar to those used in \eqref{holographically renormalized action} to renormalize the action. Indeed, one can show that the following renormalization procedure \cite{Compere:2008us , Compere:2020lrt , Fiorucci:2020xto} has the requiered properties: 
\begin{equation}
\begin{split}    \Theta^r_{ren}[g; \delta g]  =& \Theta^r_{EH}[g] + \delta   L_{GH}[g] + \delta L_{ct} [g] + \delta L_\circ[g] + \delta L_b[g] + \delta L_R [g]\\
    &- \partial_a \Theta^a_\circ[g;\delta g] - \frac{1}{2} r \partial_a \bar{\Theta}^a_\circ [g;\delta g]\,.
    \label{thetarenBondi}
    \end{split}
   \end{equation} 
   The first line is the part of the presymplectic potential prescribed by the renormalized action \eqref{holographically renormalized action}. It fixes the $\delta$-exact ambiguity in \eqref{ambiguity in theta} as $A^r =   L_{GH} +  L_{ct}  +  L_\circ + L_b + L_R$. The second line fixes the $d$-exact ambiguity appearing in \eqref{ambiguity in theta} as
   \begin{equation}
       Y^{ar}[g; \delta g] =  \Theta^a_\circ [g; \delta g]+ \frac{1}{2} r  \bar{\Theta}^a_\circ [g; \delta g]\,.
       \label{IY ambiguity explicit}
   \end{equation} Here, $\Theta^a_\circ [g; \delta g]$ is the presymplectic potential associated with the corner Lagrangian \eqref{corner lagrangian}, \textit{i.e} $\delta L_\circ = \partial_a \Theta_\circ^a$. To understand the second term in \eqref{IY ambiguity explicit}, we define the following unphysical quantities induced on the spacetime boundary:
   \begin{equation}
       \bar{v}^a = \lim_{r\to\infty} \left( r v^a \right), \qquad \bar{\gamma}_{ab} = \lim_{r\to\infty} \left( \frac{1}{r^2} {\gamma}_{ab} \right) \label{pullbacked quantities}
   \end{equation} and we write $\bar{\gamma} = \det (\bar{\gamma}_{ab})$ and $\bar{D}_a$ the determinant and the Levi-Civita connection associated with the induced boundary metric $\bar{\gamma}_{ab}$, respectively. Notice that the boundary vector $\bar{v}^a$ corresponds to the relativistic velocity of the holographic fluid in Bondi frame \cite{Ciambelli:2020eba , Ciambelli:2020ftk}. The term $\bar{\Theta}^a_\circ$ is then understood as the presymplectic potential of the boundary Lagrangian
\begin{equation}
    \bar{L}_\circ[\bar{v}^a;\bar{\gamma}_{ab}] = - \frac{1}{8 \pi G} \sqrt{-\bar{\gamma}} \bar{D}_a \bar{v}^a, \qquad  \delta \bar{L}_\circ = \partial_a \bar{\Theta}^a_\circ = - \frac{1}{8 \pi G} \sqrt{-\bar{\gamma}} \bar{D}_a \delta \bar{v}^a
\end{equation}  where the variation is taken with respect to $\bar{v}^a$ by keeping the boundary metric $\bar{\gamma}_{ab}$ fixed (\textit{i.e.} $\bar{\gamma}_{ab}$ is seen as a background). Hence, we see that the counter-terms to renormalize the presymplectic potential \eqref{thetarenBondi} are of the same nature than those necessary to remove the divergences in the action \eqref{holographically renormalized action}.

We have explicitly
\begin{equation}
\begin{split}    \Theta^r_{ren}[g; \delta g] = \frac{1}{16\pi G} &[ e^{\varphi}M\delta(\varphi - 2\beta_0) + 2e^{\varphi} N\delta U_0 \\
    & + 2 e^{2\beta_0- \varphi} (6 \partial_\phi \beta_0  \partial_\phi \delta \beta_0 - \partial_\phi \varphi \partial_\phi \delta \beta_0 + \partial_\phi^2 \delta \beta_0) ]+  \mathcal{O}(r^{-1}) \,.
    \label{presymplectic potential Bondi 3d}
    \end{split}
\end{equation} The associated presymplectic current reads as
\begin{equation}
\begin{split}
    \omega^r_{ren} [g; \delta_1 g, \delta_2 g] = \frac{1}{16\pi G} &[ \delta_2 (e^\varphi M) \delta_1 (\varphi - 2 \beta_0) + 2 \delta_2 (e^\varphi N) \delta_1 U_0 -2 e^{2 \beta_0 - \varphi} \partial_\phi \delta_2 \varphi \partial_\phi \delta_1 \beta_0 \\
    &+ 2 \delta_2 (2 \beta_0 - \varphi ) e^{2\beta_0- \varphi} (6 \partial_\phi \beta_0  \partial_\phi \delta_1 \beta_0 - \partial_\phi \varphi \partial_\phi \delta_1 \beta_0 + \partial_\phi^2 \delta_1 \beta_0) ] \\ &- (1 \leftrightarrow 2)+  \mathcal{O}(r^{-1})\,.
    \label{renormalized symplectic current}
    \end{split}
\end{equation} 
Notice that this expression vanishes at leading order when we impose Dirichlet boundary conditions \eqref{Dirichlet 3d}. Hence, the associated charges are conserved  and the variational principle \eqref{holographically renormalized action} is stationary on solutions, which is in agreement with our general discussion in section \ref{sec:Conservation and variational principle}.

\subsection{Integrability and charge algebra}
In this section, we discuss the renormalized charges and present a particular slicing of the phase space for which they are integrable.  Furthermore we compute the charge algebra.

\subsubsection{Surface charges}

Let us mention that the holographic renormalization procedure \eqref{thetarenBondi} does not affect the finite part in $r$ of the Iyer-Wald charges. Furthermore, since the Barnich-Brandt and the canonical Iyer-Wald procedures coincide in Bondi gauge (\textit{i.e.} $E^{ru}[g; \delta g , \delta g] =0$, see \eqref{E in 3d}), the finite charge expressions that we discuss now correspond to the finite part of the Barnich-Brandt charges \eqref{BB charges in 3d} as well. 

The renormalized co-dimension $2$ form can be derived using 
\begin{equation}
    \partial_a k^{ra}_{ren,\xi} [g; \delta g] = \omega^r_{ren}[g; \delta_\xi g, \delta g]
    \label{fundamental relation}
\end{equation} where $\omega^r_{ren}[g; \delta_1 g, \delta_2 g]$ is provided in \eqref{renormalized symplectic current}. As discussed in section \ref{sec:Conservation and variational principle}, this defines the co-dimension $2$ form up to a total derivative term that will not play any role when integrating on the circle $S^1_\infty$ at infinty. We obtain the infinitesimal charges by integration on $S^1_\infty$
\begin{equation}
    \ndelta{Q}_\xi[g] = \int_0^{2\pi} d\phi ~k^{ur}_{ren,\xi} [g; \delta g]\, .
\end{equation} The explicit expression reads as
\begin{align}
    \ndelta Q_\xi[g]= \frac1{8\pi G} \int_0^{2\pi} d\phi ~\Big[& Y \delta\left( e^{\varphi} N \right) + \partial_\phi \left( e^{2\beta_0-\varphi} \partial_\phi f \right) \delta\left( \beta_0-\varphi\right) \nonumber \\ 
    &+ f \Big(\frac12  e^{\varphi}\delta M - e^{\varphi} M   \delta\left( \beta_0-\varphi\right) -U_0 \delta\left( e^{\varphi} N \right) \nonumber \\
    & + e^{2\beta_0- \varphi} (6 \partial_\phi \beta_0  \partial_\phi \delta \beta_0 - \partial_\phi \varphi \partial_\phi \delta \beta_0 + \partial_\phi^2 \delta \beta_0  \Big)\Big]\,.
    \label{charges 3d v1}
\end{align} The charges are finite, thanks to the renormalization procedure \eqref{thetarenBondi}. 
As a consequence of \eqref{fundamental relation}, they are generically not conserved. Moreover, we notice that the Weyl charge associated with the parameter $\omega$ vanishes. Therefore, the Weyl rescaling part  of the residual gauge diffeomorphisms is not in the asymptotic symmetry algebra. This observation contrasts with results obtained in the Fefferman-Graham gauge \cite{Fiorucci:2020xto , Alessio:2020ioh} where the Weyl charge is non-vanishing and highlights the presence of a Weyl anomaly in the dual theory.  However, this apparent discrepancy unveils some non-trivial dependence in the choice of gauge fixing to perform the analysis of asymptotics. Indeed, the diffeomorphism between Bondi and Fefferman-Graham gauge being field-dependent \cite{Poole:2018koa , Compere:2019bua , Ciambelli:2020eba, Ciambelli:2020ftk} the symplectic structure and the associated charges transform in a subtle way \cite{Compere:2016hzt , Compere:2020lrt}. We will address the question on how the symplectic structure transforms under field-dependent diffeomorphisms elsewhere.

Finally, more importantly for us, the charges \eqref{charges 3d v1} seem to be non-integrable. As discussed in section \ref{sec:Integrability}, this apparent obstruction for integrability can be cured by performing field-dependent redefinitions of the symmetry parameters, which amounts to solve the Pfaff problem \cite{Barnich:2007bf , Adami:2020ugu ,  Compere:2017knf , Grumiller:2019fmp ,  Alessio:2020ioh, Ciambelli:2020shy}. In our case, we perform the redefinition 
    \begin{align}
   \tilde{f}=  f \, e^{2\beta_0-\varphi} , \qquad \tilde{Y}= Y - U_0\, f , \qquad \tilde{\omega} = \omega
  \label{redefinition para 3d}
\end{align} where $\tilde{f}$, $\tilde{Y}$ and $\tilde{\omega}$ are taken to be field-independent, \textit{i.e.} $\delta \tilde{f} = \delta \tilde{Y} = \delta \tilde{\omega} = 0$. In terms of these parameters, the commutation relations \eqref{commutation relations 3d} become $[\xi(\tilde{f}_1,\tilde{Y}_1, \tilde{\omega}_1), \xi(\tilde{f}_2,\tilde{Y}_2, \tilde{\omega}_2) ]_\star = \xi(\tilde{f}_{12},\tilde{Y}_{12}, \tilde{\omega}_{12})$ with
\begin{equation}
\begin{split}
     \tilde \omega_{12}&= 0 \,, \\
  \tilde f_{12}& =\tilde Y_1 \partial_\phi \tilde f_2+\tilde f_1 \partial_\phi \tilde Y_2 - (1\leftrightarrow2) \, , \\
   \tilde Y_{12}& =\tilde  Y_1 \partial_\phi \tilde Y_2+\frac1 {\ell^2} \tilde f_1 \partial_\phi \tilde f_2 - (1\leftrightarrow2)\,.
   \label{algebra of good parameters}
   \end{split}
\end{equation}
These commutation relations higlight the structure of a direct sum between the abelian Weyl rescalings $C_\infty^{(\omega)} (\mathscr{I})$ and a Lie algebroid with a one-dimensional base space parametrized by $u$ \cite{Crainic}. In the asymptotically locally flat case ($\ell \to \infty$), the algebra at each value of $u$ is the BMS$_3$ algebra given by the semi-direct sum Diff$(S^1)\loplus$Vect$(S^1)$. In the asymptotically locally AdS$_3$ case, the algebra at each value of $u$ is the double copy of the Witt algebra Diff$(S^1)\oplus$Diff$(S^1)$. The later can be made manifest by rewriting the algebra \eqref{algebra of good parameters} in terms of the parameters $F = \frac{1}{\ell} \tilde f + \tilde Y$ and $G =-\frac{1}{\ell}\tilde f +\tilde Y$ as
\begin{equation}
    \tilde \omega_{12}= 0, \quad F_{12} = F_1 \partial_\phi F_2 - F_2 \partial_\phi F_1 , \quad G_{12} = G_1 \partial_\phi G_2 - G_2 \partial_\phi G_1\,.
\end{equation}
The redefinition \eqref{redefinition para 3d} renders the charges \eqref{charges 3d v1} integrable. We have explicitly 
\noindent $\ndelta Q_\xi [g] \equiv \delta Q_\xi [g]$ with
\begin{align} 
  \delta Q_\xi [g]  = \frac1{8\pi G} \int_0^{2\pi} d\phi ~ \Big[& \tilde Y \delta\left( e^{\varphi} N \right) + \tilde f\delta \Big( \frac12 e^{2\varphi-2\beta_0}M \Big)  \nonumber    \\
    &+ \tilde f\delta \Big(4(\partial_\phi \beta_0)^2-2 \partial_\phi \beta_0\partial_\phi \varphi+\frac12 (\partial_\phi \varphi)^2 + \partial_\phi^2(2\beta_0-\varphi)  \Big) \Big]
     \label{intcharges}
\end{align} where we threw away a total derivative in $\phi$. Integrating the expression \eqref{intcharges} on a path in the solution space gives the finite charge expression
\begin{align}
    Q_\xi[g] = \frac1{16\pi G} \int_0^{2\pi} d\phi ~ \Big[& 2 \tilde Y   \tilde N  + \tilde f \tilde M \Big]
    \label{integrable charges 3d}
\end{align} where
\begin{equation}
\begin{split}
\tilde{N} = e^{\varphi} N \,, \qquad
    \tilde{M} =  e^{2\varphi-2\beta_0}M + 
    8(\partial_\phi \beta_0)^2-4 \partial_\phi \beta_0\partial_\phi \varphi+ (\partial_\phi \varphi)^2 + 2 \partial_\phi^2(2\beta_0-\varphi)\,.
    \end{split}
\end{equation}
Notice that there are only two independent charges. However, one expects from the arguments presented in \cite{Grumiller:2020vvv} that the maximal number of independent charges in this context is three. As already suggested in \cite{Ciambelli:2020eba, Ciambelli:2020ftk}, this confirms that three-dimensional Bondi gauge is not the most general framework to study the maximal phase space of the theory.

\subsubsection{Charge algebra}

The charges \eqref{integrable charges 3d} being integrable, they form a representation of the symmetry algebra, up to a possible central extension \cite{Brown:1986nw , Brown:1986ed , Barnich:1991tc, Barnich:2001jy, Barnich:2007bf}. Indeed, using the bracket
\begin{equation}
    \{ Q_{\xi_1}[g], Q_{\xi_2}[g] \} \equiv \delta_{\xi_2} Q_{\xi_1}[g] \,,
\end{equation} we obtain
\begin{equation}
    \{ Q_{\xi_1}[g], Q_{\xi_2}[g] \} = Q_{[\xi_1 , \xi_2]_\star} [g]+  \frac1{8\pi G}  \int_0^{2\pi} d\phi ~\left( \partial_\phi^2 \tilde f_1 \, \partial_\phi \tilde Y_2 -\partial_\phi^2 \tilde f_2 \, \partial_\phi \tilde Y_1  \right)\, \label{charge algebra 3d}
\end{equation}
where ${[\xi_1 , \xi_2]_\star} $ is given by \eqref{algebra of good parameters}. Let us stress that the charge algebra is $u$-dependent. In other words, it corresponds to a centrally extended Lie algebroid whose one-dimensional base space is parametrized by the boundary time $u$. The centrally extended algebra at each value of $u$ is the double copy of the Virasoro algebra in the asymptotically locally AdS$_3$ case, and the centrally extended BMS$_3$ algebra in the asymptotically locally flat case.   

We now restrict our general charge algebra for the particular case of Dirichlet boundary conditions \eqref{Dirichlet 3d}. Defining 
\begin{equation}
   \tilde f = \frac{\ell}{2} (Y^+ + Y^-), \qquad \tilde Y = \frac{1}{2} (Y^+ -Y^-) 
\end{equation} and $x^\pm = \frac{t}{\ell}\pm \phi$, the conformal Killing equations \eqref{ckv 3d} become simply $\partial_\pm Y^\mp = 0$. The commutation relations \eqref{algebra of good parameters} are $Y^\pm = Y^\pm_1 \partial_\pm  Y^\pm_2 - Y^\pm_1 \partial_\pm  Y^\pm_1$, which is precisely the direct sum of two copies of the Witt algebra. 
 Moreover, using the decomposition in modes $Y^\pm = \sum_{m\in \mathbb{Z}} Y_m^\pm l_m^\pm$, $l_m^\pm = e^{\pm i m x^\pm}$ and writing $L^\pm = Q_{\xi(l_m^\pm)}$, the charge algebra \eqref{charge algebra 3d} is
\begin{equation}
  i  \{ L_m^\pm, L_n^\pm \} = (m-n) L_{m+n}^\pm - \frac{c^\pm}{12} m^3 \delta^0_{m+n} , \qquad \{ L_m^\pm, L_n^\mp \} = 0
  \label{BH central}
\end{equation} where $c^\pm = \frac{3\ell}{2G}$ is precisely the Brown-Henneaux central charge \cite{Brown:1986nw}.

\subsection{Flat limit}
We now  discuss the flat limit of our three-dimensional results that goes from asymptotically locally AdS$_3$ to asymptotically locally flat spacetimes. The Bondi gauge has the interesting property to have a well-defined behaviour when $\ell \to \infty$ \cite{Barnich:2012aw} . 

The flat limit of the solution space and of the residual gauge diffeomorphisms is discussed in details in \cite{Ciambelli:2020eba , Ciambelli:2020ftk}. Let us mention that the boundary metric \eqref{induced boundary metric} becomes degenerate when $\ell \to \infty$, so that the timelike spacetime boundary ($\mathscr{I}_{AdS}$) becomes null in the limit ($\mathscr{I}^+$). As mentioned under \eqref{properties va}, the degeneracy direction on the boundary is generated by the pull-back on the boundary of the vector $v^a$ introduced in \eqref{corner lagrangian}. Moreover, the symmetry algebra of the residual gauge symmetries in the integrable slicing is discussed below \eqref{algebra of good parameters}.

Now, at the level of the phase space, our analysis yields additional results. Indeed, we have seen that the on-shell value of the renormalized action was given by \eqref{on-shell action}. This expression being finite in $\ell$, its flat limit is straightforward. Its Euclidean version provides us with the expression for the free energy in asymptotically locally flat spacetimes. Similarly, the renormalized presymplectic potential \eqref{presymplectic potential Bondi 3d} and presymplectic current \eqref{renormalized symplectic current} do not depend on $\ell$ and their flat limit can be readily taken. Finally, the charges \eqref{integrable charges 3d} and the algebra \eqref{charge algebra 3d} do not depend on $\ell$ and the expressions are therefore formally the same in asymptotically locally flat spacetimes.

Notice that even if the phase space expressions appear to be the same in asymptotically locally AdS$_3$ and asymptotically locally flat spacetimes, the physics behind it is completely different. Indeed, as discussed in \cite{Ciambelli:2020eba , Ciambelli:2020ftk}, the constraint equations on $M$ and $N$ are not the same in the flat limit. Analogously, the variation of the solution space is $\ell$-dependent. To understand the implications of that, consider the boundary conditions that define asymptotically flat spacetimes \cite{Barnich:2006av , Ashtekar:1996cd}. In our framework, these conditions have formally the same form as the Dirichlet boundary conditions \eqref{Dirichlet 3d} in AdS. However, the constraint equations on the parameters \eqref{ckv 3d} reduce to 
\begin{equation}
    \partial_u f = \partial_\phi Y , \qquad \partial_u Y = 0,\qquad \omega = 0 \label{ckv 3d flat}
\end{equation} in the limit $\ell \to \infty$ \cite{Barnich:2012aw}. This can be readily solved as $Y = Y(\phi)$ and $
    f = T(\phi) + u \, \partial_\phi Y$,
 where $T(\phi)$ are the supertranslation generators, while $Y(\phi)$ are the superrotation generators. 
Using the mode decomposition $T_n = \xi (T = e^{in\phi}, Y=0)$ and $Y_n = \xi (T = 0, Y= e^ {i n \phi })$ and writing $P_n = Q_{T_n}[g]$ and $J_n = Q_{Y_n}[g]$,  the charge algebra \eqref{charge algebra 3d} reads as
\begin{equation}
    [P_m , P_n] = 0, \quad  i[J_m, J_n] = (m-n) J_{m+n}, \quad i[J_m , P_n] = (m-n)P_{m+n} - \frac{m^3}{4\pi G} \delta_{m+n}^0\,.
    \label{BMS central charge}
\end{equation} This is  the BMS$_3$ centrally extended algebra  \cite{Barnich:2006av}. Hence, when fixing the boundary structure, we see that the central charge appearing in \eqref{charge algebra 3d} reduces to the Brown-Henneaux central extension \cite{Brown:1986nw} in asymptotically AdS$_3$ spacetimes (see equation \eqref{BH central}) and to the BMS$_3$ central extension \cite{Barnich:2006av} in the flat limit $\ell \to \infty$ (see equation \eqref{BMS central charge}).

\section*{Comments}
\addcontentsline{toc}{section}{Comments}

In this section, we provide further comments on the results that have been obtained in this work and draw some motivations for this analysis in a broader context.

In this article, we have investigated a type of boundary conditions that allows the boundary structure to fluctuate. The systematic analysis of such boundary conditions is recent. In asymptotically locally flat spacetimes, the generalized BMS group, which is an infinite-dimensional enhancement of the BMS group with smooth superrotations, has been obtained by relaxing the conditions on the transverse components of the boundary metric \cite{Campiglia:2014yka, Campiglia:2015yka , Compere:2018ylh}. Similar relaxed boundary conditions have been considered in presence of non-vanishing cosmological constant, leading to the $\Lambda$-BMS group(oid) \cite{Compere:2019bua , Compere:2020lrt}. These additional symmetries are believed to play an important role in the relations between asymptotic symmetries, soft theorems and memory effects (see \textit{e.g.} \cite{Strominger:2017zoo} for a review). 

In asymptotically locally AdS spacetimes, as explained in \cite{Fiorucci:2020xto}, the boundary conditions with relaxed boundary structure (also referred as ``leaky boundary conditions'') yield some non-vanishing symplectic flux throughout the boundary. In this picture, the gravitational system is seen as an open system which couples to external sources encoded by the fluctuations of the boundary metric \cite{Troessaert:2015nia , Wieland:2020gno}. The inclusion of the boundary structure in the phase space leads to non-conservation and non-integrability of charges which are not necessarily related to the passage of gravitational waves through the spacetime boundary. In particular, as we have seen in the present context, it also produces non-conservation and non-integrability in lower-dimensional gravity theories. Understanding the precise role of these sources in the AdS/CFT correspondence would be a rewarding investigation. 

In particular, using flat limit processes similar to those that have been studied here, a description of holography with external sources would enable us to capture some features of holography in higher-dimensional radiating asymptotically flat spacetimes. A first step towards this endeavour is the holographic renormalization procedure that we have performed in the present paper in the Bondi gauge. We have found the covariant counter-terms to remove the potential divergences when taking the cut-off to infinity or when taking the flat limit. We believe that this procedure could be repeated in higher dimension without conceptual obstruction and ultimately lead to a renormalized action in asymptotically flat spacetimes at null infinity.

This work provides a non-trivial check of the conjecture coined in \cite{Adami:2020ugu} and stating that non-integrability, when not due to genuine propagating degrees of freedom, can be removed by a field-dependent redefinition of the symmetry parameters.  We have shown that for asymptotic boundaries such redefinitions \eqref{new parameters} and \eqref{redefinition para 3d} render the charges integrable. Starting from any integrable slicing, one can generate an infinite number of integrable slicings that lead to different symmetry algebras or algebroids, see the general discussion in  \cite{Adami:2020ugu}. In our case, we have chosen slicings that display pleasant properties. Indeed, they lead to some field-independent commutation relations (see equations \eqref{symmetry algebra 2d} and \eqref{algebra of good parameters}) and yield some simple transformation laws for the solution space (see \textit{e.g.} \eqref{simple variations 2d}). Furthermore, as discussed in the text, they have the property to be compatible with the restriction to Dirichlet boundary conditions, which amounts to freeze the boundary structure and turn off the sources. 
It would be interesting to see whether the procedures to render the charges integrable can be applied to higher dimensions when non-integrability that is not related to propagating degrees of freedom occurs (\textit{e.g.} in the context of asymptotically flat spacetimes at spatial infinity, where there is no flux of gravitational waves, but where relaxing the boundary structure is possible).


\section*{Acknowledgements}

We are indebted to Daniel Grumiller and Shahin Sheikh-Jabbari for crucial discussions and comments on the manuscript. We also warmly thank Hamed Adami, Glenn Barnich, Luca Ciambelli, Geoffrey Comp\`ere, Adrien Fiorucci, Charles Marteau, Marios Petropoulos, Vahid Taghiloo and Hossein Yavartanoo for their precious comments and collaborations on related subjects. RR was supported by the Austrian Science Fund (FWF), project P 32581-N. CZ was supported by the Austrian Science Fund (FWF), projects P 30822 and M 2665.

\appendix

\section{Solution space in linear dilaton Bondi gauge}\label{app2d}

In this appendix we discuss the solution phase for the theory \eqref{DGT} (we keep $U$ and $V$ arbitrary). We only assume that the the metric is in Bondi gauge
\begin{equation}
    ds^2 = 2 B(u,r) du^2 - 2 e^{A(u,r)} du dr\,.
\end{equation}

We start by analysing the $rr$ component of \eqref{GMeomG}, it reads as 
\begin{equation} \label{GMeomGrr}
    \partial_r^2 X+\partial_r X(U \partial_r X-\partial_rA)=0\,.
\end{equation}
We have two classes of solutions, the constant dilaton and non-constant dilaton. The former yields to $ V(X) =0$. Indeed taking the trace of \eqref{GMeomG} yields \begin{equation}\label{GMtraceemo}
    \nabla^2 X= -2V\,.
\end{equation}

The non-constant dilaton case is more interesting \cite{Grumiller:2015vaa} and for now on we suppose that $\partial_r X\neq0$. Equation \eqref{GMeomGrr} can be solved by
\begin{equation} \label{coordradial}
    \partial_r X= e^{-Q(X)+A(u,r)}\,,\qquad  Q(X):=q_0(u)+\int^X dY U(Y)\,.
\end{equation}
Hence \eqref{GMtraceemo} reads as 
\begin{equation}
    \partial_rX\partial_r B=e^{2A}V+BU(\partial_rX)^2-e^A\partial_r\partial_u X
\end{equation}
which can be solved by
\begin{equation}\label{Beq}
B=e^{Q(X)} \left(B_0+\int^Xd Y(V(Y)e^{Q(Y)} + \partial_u (Q(Y)-A(Y) ) \right)\,.
\end{equation}
This solves all equations of motion but the evolution equation for $X$, namely
\begin{equation}\label{app2devoleq}
    \partial_u^2X+U\, (\pu  X )^2 -\pu X\, \pu A +\pu B_0 -B\,e^{-Q}\, \pu q_0=0\,.
\end{equation}

Furthermore, we only have used one gauge degree of freedom to fix $g_{rr}=0$. We can use the other one to require the dilaton to be linear, namely $\partial_r e^{-Q+A}=0$. It implies $A=Q(X)-Q_0(u)$ with $Q_0(u)$ an arbitrary function and
\begin{equation}
    X=e^{-Q_0(u)}r+\jf_0(u)\,.
\end{equation}
Note that in more standard Eddington-Finkelstein type of gauges, where $A =0$, the linear dilaton condition enforces $U=0$ and $q_0=Q_0$. 

\addcontentsline{toc}{section}{References}

\providecommand{\href}[2]{#2}\begingroup\raggedright\endgroup

\end{document}